\documentclass[12pt]{article}
\usepackage{amsmath,amssymb,theorem,cite,epsfig,url,psfrag,eepic,amsmath,mathtools}
\mathtoolsset{showonlyrefs}
\usepackage[hidelinks]{hyperref}
\advance \topmargin by -\headheight
\advance \topmargin by -\headsep
\evensidemargin \oddsidemargin
\numberwithin{equation}{section}

\makeatletter
\def\Appendix{\appendix
  \def\@seccntformat##1{Appendix~\csname the##1\endcsname.~~}}
\makeatother

\def\XXint#1#2#3{{\setbox0=\hbox{$#1{#2#3}{\int}$}
\vcenter{\hbox{$#2#3$}}\kern-.5\wd0}}

\usepackage{dynkin-diagrams}

\DeclareMathOperator{\Tr}{Tr}
\DeclareMathOperator{\End}{End}
\DeclareMathOperator{\Lie}{Lie}

\newcommand{\SU}{\mathrm{SU}}
\newcommand{\SO}{\mathrm{SO}}
\newcommand{\GL}{\mathrm{GL}}
\newcommand{\SL}{\mathrm{SL}}
\begin{document}
\title{\textbf{Yang--Baxter deformed sigma models on $\SU(N)/\SO(N)$}\vspace*{0.3cm}}
\date{}
\author{Danila Alexandrov$^{1,2}$, Alexey Bychkov$^{1,2,3}$, Ivan Kalinichenko$^{1,4}$ and Alexey Litvinov$^{1}$
\\[\medskipamount]
\parbox[t]{0.85\textwidth}{\normalsize\it\centerline{1. Skolkovo Institute of Science and Technology, 121205 Moscow, Russia}}
\\
\parbox[t]{0.85\textwidth}{\normalsize\it\centerline{2. Moscow Institute of Physics and Technology, 141700 Dolgoprudny, Russia}}
\\
\parbox[t]{0.85\textwidth}{\normalsize\it\centerline{3. HSE University, 6 Usacheva str., Moscow 119048, Russia}}
\\
\parbox[t]{0.85\textwidth}{\normalsize\it\centerline{4. Institute for Theoretical and Mathematical Physics MSU,  119991 Moscow, Russia}}}
\maketitle
\begin{abstract}
    We study Yang--Baxter deformations of two-dimensional sigma models on compact symmetric spaces $\SU(N)/\SO(N)$, with particular emphasis on their ultraviolet behavior. We claim that the ultraviolet fixed point is described by the $\SO(N,\mathbb C)/\SO(N,\mathbb R)$ non-compact WZW model together with $N-1$ compact free bosons. We determine the leading correction to this fixed point and identify it with the perturbation generated by the corresponding screening currents. We also extend our construction to the broader class of symmetric spaces of maximal rank.
\end{abstract}
\section{Introduction}
Two-dimensional nonlinear sigma models \cite{GellMann:1960np} provide a rich framework in which geometry, renormalization group flow, and integrability (both classical and quantum) can be explored simultaneously. When the target space is a symmetric manifold $G/H$, the classical equations of motion admit a zero-curvature representation \cite{Eichenherr:1979ci}, while in a number of cases integrability survives quantization and leads to an exact factorized scattering theory \cite{Zamolodchikov:1978xm} in the weakly coupled region. The compact symmetric space
\begin{equation}
\frac{\SU(N)}{\SO(N)}
\end{equation}
is one of the basic examples in this class. The corresponding sigma model is asymptotically free and belongs to one of the standard families of exactly solvable QFTs. Namely, many physical properties, such as the particle spectrum, the $S$-matrix, the ground-state energy, etc., can be found exactly using thermodynamic Bethe ansatz methods. The structure of the symmetric pair $(\SU(N),\SO(N))$ plays a central role in these considerations (see, for example, \cite{Fendley:2001hc} and the references therein).

A broad class of integrability-preserving deformations of sigma models is given by the Yang--Baxter construction. Initially introduced for principal chiral models \cite{Klimcik:2008eq} and subsequently extended to sigma models on symmetric spaces \cite{Delduc:2013fga}, the Yang--Baxter deformation (also conveniently called the $\eta$-deformation) is determined by a solution of the modified classical Yang--Baxter equation \cite{Belavin1982}, called the $\mathcal{R}$-operator, which explicitly enters the classical action. This deformation changes the target space metric and, in general, generates a non-trivial $B$-field, while preserving Lax integrability. At the same time, the manifest global symmetry of the undeformed theory is replaced by a symmetry of the quantum group (see \cite{Hoare:2021dix} for a review). These properties make Yang--Baxter deformations natural candidates for integrable quantum field theory. They also raise a basic question: what conformal field theory describes the short-distance behavior of the deformed model, and which operators govern the vicinity of this ultraviolet fixed point?

A useful approach to this question was developed in \cite{Fateev:1996ea,Fateev:2017mug,Fateev:2018yos,Litvinov:2018bou,Fateev:2019xuq,Litvinov:2019rlv,Alfimov:2020jpy}. In this approach, an integrable QFT is characterized near its ultraviolet fixed point by a system of screening charges. The commutant of these charges determines a chiral algebra and an infinite family of local integrals of motion. For the Yang--Baxter deformations studied previously, this construction often leads to a Gaussian ultraviolet theory with a number of compact and non-compact bosons perturbed by exponential ``graviton'' operators of Wakimoto type \cite{Wakimoto:1986gf,zbMATH04093704}. It was also noticed that the same integrable structure admits a dual Toda-like description\footnote{ Such a dual description for $\SO(3)/\SO(2)$ sausage model was first suggested by Alexey Zamolodchikov in his unpublished paper.}. This picture suggests that the screening system is more fundamental than any particular Lagrangian realization.

In the present paper, we apply this philosophy to the Yang--Baxter deformation of the $\SU(N)/\SO(N)$ sigma model. The ultraviolet behavior of this family turns out to be qualitatively different from the purely Gaussian theories encountered in the previous examples. Our main result is the identification
\begin{equation}\label{UV-conjecture}
\frac{\SU(N)}{\SO(N)}\Bigg|_\text{UV fixed point}
\quad\simeq\quad
\left(\frac{\SO(N,\mathbb C)}{\SO(N,\mathbb R)}\,\textrm{ gauged WZW}\right)\quad
\times\quad\mathrm{U}(1)^{N-1}\, .
\end{equation}
Here, the first factor is the non-compact conformal sigma model \cite{Gawedzki:1991yu} associated with the symmetric space $\SO(N,\mathbb C)/\SO(N,\mathbb R)$, whereas $\mathrm{U}(1)^{N-1}$ denotes $N-1$ free compact bosons. Thus, the Yang--Baxter deformation does not simply flatten the compact target space in the ultraviolet. Instead, its degrees of freedom reorganize into an interacting non-compact orthogonal sector and a compact abelian sector of rank $(N-1)$. 

The decomposition \eqref{UV-conjecture} admits a natural interpretation in terms of the deformed geometry of the compact manifold $\SU(N)/\SO(N)$ in the UV limit. The Cartan directions of $\SU(N)$ give rise to the compact $\mathrm{U}(1)^{N-1}$ abelian (non-interacting) sector, while the remaining fields assemble into the non-compact $\SO(N,\mathbb{C})/\SO(N,\mathbb{R})$ sector. Together, these fields provide natural coordinates on the stratum obtained by intersecting the Cartan embedding of $\SU(N)/\SO(N)$ with the large Bruhat-type cell of $\mathrm{GL}(N,\mathbb C)$ \cite{BagnoCherniavsky2012,RichardsonSpringer1990}.

In addition, we determine the first correction to the fixed point \eqref{UV-conjecture}. Schematically, it has the form
\begin{equation}\label{UV-expansion}
S_{\mathrm{YB}}=S_{\mathrm{UV}}
 +\lambda\sum_{r=1}^{N-1}\int d^{2}z\,
\mathcal O_{r}(z,\bar z)
+\dots,
\end{equation}
where $\lambda$ is the running coupling that vanishes in the ultraviolet. 

The paper is organized as follows. In Section~\ref{YB-model} we describe the Yang--Baxter deformation of symmetric space sigma models and provide details for the cases of $\SU(2)/\SO(2)$ and $\SU(3)/\SO(3)$. In Section~\ref{UV-theory} we study its renormalization group flow and extract the ultraviolet asymptotics, which we identify with $\SO(N,\mathbb C)/\SO(N,\mathbb R)\times \mathrm{U}(1)^{N-1}$ CFT. We also derive the leading UV correction. In Section~\ref{screenings} we discuss quantization of $\SU(N)/\SO(N)$ YB sigma model within conformal perturbation theory. We match the first correction  with the screening perturbation. In Section ~\ref{geometry} we analyze the stratification of the geometry of $\SU(N)/\SO(N)$. In Section~\ref{generalizations}, we extend our construction to other symmetric spaces of maximal rank. In Section~\ref{concl} we present our concluding remarks and formulate several open problems.
\section{YB-deformed sigma models on symmetric spaces}\label{YB-model}
We begin by recalling the Yang--Baxter deformation of a sigma model on a compact symmetric space (for more details, see \cite{Hoare:2021dix}). 

Let $G$ be a compact simple Lie group, and let $\sigma:G\longrightarrow G$ be an involutive automorphism. The closed subgroup fixed by $\sigma$ will be denoted by $H$. Denote the corresponding Lie algebras by $\mathfrak g = \Lie(G)$ and $\mathfrak h =\Lie(H)$. The differential of $\sigma$ at the identity gives the symmetric decomposition:
\begin{equation}\label{z2-grading}
  \mathfrak g=\mathfrak g^{(0)}\oplus\mathfrak g^{(1)},
  \qquad
  \mathfrak g^{(0)}=\mathfrak h,
\end{equation}
where $\sigma$ (or, rather, its differential at $e\in G$) acts with the eigenvalues $+1$ and $-1$ on $\mathfrak g^{(0)}$ and $\mathfrak g^{(1)}$, respectively.  In particular,
\begin{equation}
  [\mathfrak g^{(0)},\mathfrak g^{(0)}]\subseteq\mathfrak g^{(0)},
  \qquad
  [\mathfrak g^{(0)},\mathfrak g^{(1)}]\subseteq\mathfrak g^{(1)},
  \qquad
  [\mathfrak g^{(1)},\mathfrak g^{(1)}]\subseteq\mathfrak g^{(0)}.
  \label{eq:symmetric-space-brackets}
\end{equation}
We write $P_0$ and $P_1$ for the corresponding projectors.

The additional ingredient used in the Yang--Baxter construction is the $\mathcal{R}$-operator $\mathcal{R} \in \End(\mathfrak g)$. It is required to be a skew-symmetric non-split solution of the modified classical Yang--Baxter equation, i.e. it satisfies
\begin{equation}\label{R-skew-and-MCYBE}
    \begin{aligned}
        &\Tr(\mathcal{R}(X)\cdot Y) + \Tr(X\cdot\mathcal{R}(Y)) = 0,\\
        &[\mathcal{R}X, \mathcal{R}Y] -\mathcal{R}\big([\mathcal{R}X, Y] + [ X, \mathcal{R}Y]\big) = [X, Y],
    \end{aligned}
    \qquad \forall X, Y\in\mathfrak{g}.
\end{equation}
We use the standard Drinfeld--Jimbo solution. In the Cartan--Weyl basis in $\mathfrak{g}^\mathbb{C} = \langle E_{\alpha},E_{-\alpha},H_i\rangle$ it has the form
\begin{equation}\label{DJ-matrix}
    \mathcal{R}(E_{\alpha})=iE_{\alpha},\quad
    \mathcal{R}(E_{-\alpha})=-iE_{-\alpha},\quad
    \mathcal{R}(H_i)=0.
\end{equation}

Now, let $g(x)$ be a $G$-valued field on some two-dimensional spacetime (worldsheet). Then the YB-deformed action on the symmetric space $G/H$ has the form \cite{Delduc:2013fga}
\begin{equation}\label{Coset-action-deformed}
   \mathcal{S}=-\frac{\kappa}{\hbar}\int\textrm{Tr}
\left(\left(g^{-1}\partial_{+}g\right)^{(1)}\,\frac{1}{1+i\kappa\mathcal{R}_{g}\circ\mathrm{P}_{1}}\,
   \left(g^{-1}\partial_{-}g\right)^{(1)}\right) d^{2}x,
\end{equation}
where 
\begin{equation}\label{R_g-definition}
    \mathcal{R}_{g}=\textrm{Ad}\, g^{-1}\circ\mathcal{R}\circ\textrm{Ad}\,g.    
\end{equation}
We  take the usual deformation parameter to be purely imaginary $\eta=i\kappa$. In this regime, the model exhibits asymptotic freedom.

Geometrically, the action \eqref{Coset-action-deformed} defines a metric and a $B$-field on the target space $G/H$. The metric is read off as the symmetric part of the expression \eqref{Coset-action-deformed}, and the $B$-field is the skew-symmetric one. Since the operator $\mathcal{R}\in\End(\mathfrak g^{\mathbb{C}})$ is skew-symmetric, so is $\mathcal{R}_g\in\End(\mathfrak{g}^{\mathbb{C}})$, and one explicitly has
\begin{equation}\label{deformed-g-B}
\begin{aligned}
    ds^2&=-\frac{1}{2}\Tr
\left(\left(g^{-1}dg\right)^{(1)}\,\frac{\kappa}{1+\kappa^2\left(\mathcal{R}_{g}\circ\mathrm{P}_{1}\right)^2}\,
   \left(g^{-1}dg\right)^{(1)}\right),\\
   B&=-\frac{1}{2}\Tr
\left(\left(g^{-1}dg\right)^{(1)}\,\frac{-i\kappa^2\mathcal{R}_{g}\circ\mathrm{P}_{1}}{1+\kappa^2\left(\mathcal{R}_{g}\circ\mathrm{P}_{1}\right)^2}
\wedge
\left(g^{-1}dg\right)^{(1)}\right).
\end{aligned}
\end{equation}

The model defined by \eqref{Coset-action-deformed} possesses several notable properties. In particular,
\begin{itemize}
    \item This model is classically integrable \cite{Delduc:2013fga}. In fact, the skew-symmetry and the modified classical Yang--Baxter equation \eqref{R-skew-and-MCYBE} are sufficient for the classical integrability of the underlying non-deformed coset model \cite{Eichenherr:1979ci} to survive under the deformation.
    \item What is even more remarkable is that after the deformation the model remains renormalizable at one loop \cite{Sfetsos:2015nya,Klimcik:2019kkf}. The renormalization group equation for the coupling $\kappa$ has the form
    \begin{equation}\label{RG-equation}
    \frac{d\kappa}{dt}=h_G^{\vee}\hbar\big(\kappa^2-1\big),
    \end{equation}
    where $h_G^{\vee}$ is the dual Coxeter number of $G$. The one-loop equation is solved by
    \begin{equation}
        \kappa(t)=-\tanh\left(h_G^{\vee}\hbar t\right).
    \end{equation}
    In the ultraviolet limit $t\rightarrow-\infty$, we have $\kappa\rightarrow1$.
\end{itemize}
The ultraviolet point $\kappa\rightarrow 1$ is the most interesting to us. On general physical grounds it can be argued that in the vicinity of this UV fixed point the theory admits the expansion \eqref{UV-expansion}.  

Moreover, it is believed \cite{Abdalla:1982yd} that if the ``denominator'' in $G/H$ is simple, then the undeformed model possesses the property of quantum integrability\footnote{
The simplicity of $H$ seems to be sufficient for the integrability, but not necessary; for instance, in \cite{Fendley:2001hc, Evans:2004ur} the $\mathrm{SO}(2N)/(\mathrm{SO}(N)\times\mathrm{SO}(N))$ and $\mathrm{Sp}(2N)/(\mathrm{Sp}(N)\times\mathrm{Sp}(N))$ models were shown to be quantum integrable as well.
}. It is natural to assume that the corresponding YB-deformed model \eqref{Coset-action-deformed} is also quantum integrable. It is important to emphasize that \eqref{Coset-action-deformed} is only a one-loop action and higher-order $\hbar$ corrections are needed to ensure both renormalizability and quantum integrability. The first corrections in $\hbar$ were studied in \cite{Hoare:2019ark,Hoare:2019mcc,Alfimov:2021sir}. This idea was developed in \cite{Fateev:1996ea,Fateev:2017mug,Fateev:2018yos,Litvinov:2018bou,Fateev:2019xuq,Litvinov:2019rlv,Alfimov:2020jpy,Bychkov:2025ftt,Bychkov:2025vex}. Assuming quantum integrability, the perturbing operators $\mathcal{O}_r(z,\bar{z})$ in \eqref{UV-expansion} must correspond to integrable perturbations within the conformal perturbation theory \cite{Zamolodchikov:1989zs}. Under holomorphic–antiholomorphic factorization, the holomorphic part of $\mathcal{O}_r(z,\bar{z})$ defines a screening charge. The common commutant of these charges determines a chiral algebra and an infinite family of local integrals of motion.

In this paper, we study the $\SU(N)/\SO(N)$ model. We first consider the simplest nontrivial examples.
\paragraph{An example: $\SU(2)/\SO(2)$ model.}
For a warm-up, we start with the case of $N = 2$. Recall the general form of the $\eta$-deformed action \eqref{Coset-action-deformed}. As a basis in $\mathfrak{su}(2)$ we take the following set of skew-Hermitian generators:
\begin{equation}
    t_1=\begin{pmatrix}
        0&i\\
        i&0
    \end{pmatrix},\quad
    t_2=\begin{pmatrix}
        i&0\\
        0&-i
    \end{pmatrix}\quad\text{and}\quad
    t_3=\begin{pmatrix}
        0&1\\
        -1&0
    \end{pmatrix}.
\end{equation}
We take the following parameterization of the coset element ($t_3$ generates the $\mathfrak{so}(2)$ subalgebra):
\begin{equation}\label{SU(2)-parametrization}
    g(\theta,\varphi) = \exp\left(\frac{\varphi}{2}t_2\right)\,\exp\left(\frac{\theta}{2}t_1\right),\qquad
    \theta\in \left[-\tfrac{\pi}{2},\tfrac{\pi}{2}\right],\quad\varphi\in [0,2\pi).
\end{equation}
Plugging it into \eqref{Coset-action-deformed}, we find the metric and $B$-field of the sausage model \cite{Fateev:1992tk}
\begin{equation}\label{sausage-metric-B-field}
    ds^2=\frac{\kappa}{4}\frac{d\theta^2+\cos^2\theta d\varphi^2}{1-\kappa^2\sin^2\theta},\quad
    B=\frac{i\kappa^2}{8}
    \frac{\sin2\theta}{1-\kappa^2\sin^2\theta}d\theta\wedge d\varphi.
\end{equation}
Since our target manifold is two-dimensional, the $B$-field is always a closed form. So, (at least, locally) it is exact, and thus can be neglected. For the UV expansion it is more convenient to use the coordinate $\rho$ related to $\theta$ by
\begin{equation}
    \sin\theta=\tanh\rho,\quad\rho\in [-\infty,\infty].
\end{equation}
In the coordinates $(\rho,\varphi)$ the metric in \eqref{sausage-metric-B-field} admits the expansion
\begin{equation}\label{SU(2)/SO(2)-metric-UV-expansion}
    4ds^2=d\rho^2+d\varphi^2+(\kappa-1)\cosh 2\rho\left(d\rho^2+d\varphi^2\right)+\dots
\end{equation}

The UV expansion \eqref{SU(2)/SO(2)-metric-UV-expansion} matches perfectly the following geometric realization of the target space $\SU(2)/\SO(2) \simeq S^2$. One represents $S^2$ as a disjoint union of three strata: a cylinder and two points
\begin{equation}
    S^2=\{\rho\in (-\infty,\infty),\,\varphi\in[0,2\pi]\}
    \sqcup\{\rho=\infty\}\sqcup\{\rho=-\infty\}.
\end{equation}
The leading UV metric in \eqref{SU(2)/SO(2)-metric-UV-expansion} is just the standard flat metric on the cylinder, whereas the first-order correction can be split as
\begin{equation}
    \frac{\kappa-1}{2}e^{\pm 2\rho}\left(d\rho^2+d\varphi^2\right)
\end{equation}
\paragraph{An example: $\SU(3)/\SO(3)$ model.}
Now, we consider the action \eqref{Coset-action-deformed} for $N=3$. We use the  following parameterization of the coset element:
\begin{equation}\label{SU(3)-coset-parametrization}
g=e^{\varphi_1 h_1+\varphi_2 h_2}
\begin{pmatrix}
\frac{i \sinh \left(\tau+i\phi\right)}{\sqrt{\cosh^2\rho\cosh2\tau+\sinh ^2\rho\cos2\phi}}&-\frac{i\sqrt{\cosh2\rho} \cosh\left(\tau+i\phi\right)}{\sqrt{\cosh ^2\rho\cosh2\tau+\sinh^2\rho\cos2\phi}}&0\\
\frac{\sqrt{\cosh2\rho} \cosh \left(\tau-i\phi\right)}{\sqrt{\cosh^2\rho\cosh2\tau+\sinh^2\rho\cos2\phi}}&
\frac{\sinh\left(\tau-i\phi\right)}{\sqrt{\cosh^2\rho\cosh2\tau+
\sinh^2\rho\cos2\phi}}&0\\
0&0&-i
\end{pmatrix}
\begin{pmatrix}
1&0&0\\
0&\frac{\sinh\rho}{\sqrt{\cosh2\rho}}&\frac{i\cosh\rho}{\sqrt{\cosh2\rho}}\\
0&\frac{i\cosh\rho}{\sqrt{\cosh2\rho}}&\frac{\sinh\rho}{\sqrt{\cosh2\rho}},
\end{pmatrix}
\end{equation}
where
\begin{equation}
    h_1=\begin{pmatrix}
i&0&0\\
0&-i&0\\
0&0&0
\end{pmatrix},\quad
h_2=\begin{pmatrix}
0&0&0\\
0&i&0\\
0&0&-i
\end{pmatrix}
\end{equation}
are skew-Hermitian Cartan generators of $\mathfrak{su}(3)$.
A direct calculation shows that exactly at the UV point $\kappa=1$ the deformed metric and $B$-field (up to an exact form) coincide with those of the  $\mathrm{U}(1)^{2}\times\mathbb{H}_3^+$
\begin{equation}
    \begin{aligned}
        ds^2&\Big|_{\kappa = 1}=
        \underbrace{d\varphi_1^2-d\varphi_1d\varphi_2+d\varphi_2^2}_{\mathrm{U}(1)^2}+
        \underbrace{d\rho^2+
    \cosh^2\rho\,d\tau^2+\sinh^2\rho\,d\phi^2}_{\mathbb{H}_3^+},\\
        B&\Big|_{\kappa = 1}=\frac{i}{2}\cosh2\rho\,d\tau\wedge d\phi.
    \end{aligned}
\end{equation}
Here $\mathbb{H}_3^+$ denotes the three-dimensional hyperbolic space.

The coordinates $(\rho,\tau,\phi)$ used in \eqref{SU(3)-coset-parametrization}  parameterize the hyperboloid $\mathbb{H}_3^+$
\begin{equation}
x_0^2-x_1^2-x_2^2-x_3^2=1,\qquad x_0>0,
\end{equation}
as
\begin{equation}
x_0=\cosh\rho\,\cosh\tau,\quad
x_1=\sinh\rho\,\cos\phi,\quad
x_2=\sinh\rho\,\sin\phi,\quad
x_3=\cosh\rho\,\sinh\tau,
\end{equation}
with the ranges
\begin{equation}\label{H3-ranges}
\rho\in[0,\infty),\qquad
\tau\in(-\infty,\infty),\qquad
\phi\in[0,2\pi).
\end{equation}
The Cartan angles $\varphi_{1,2}$ of $\mathrm{U}(1)^2$ naturally belong to 
\begin{equation}\label{Cartan-ranges}
    \varphi_1\in[0,2\pi),\quad
    \varphi_2\in[0,2\pi).
\end{equation}
In fact (see Section \ref{UV-theory} for more details), the coordinate ranges \eqref{H3-ranges} and \eqref{Cartan-ranges} define a four-sheeted cover of the UV manifold, which is the quotient of $\mathrm{U}(1)^{2}\times\mathbb{H}_3^+$ by the discrete group $\mathbb{Z}_2^2$, whose generators $s_{1,2}$ act as follows:
\begin{equation}\label{Z2-Z2-action}
\begin{aligned}
    &s_1:\,\varphi_1\mapsto\varphi_1+\pi,\quad
    (x_1,x_2)\mapsto(-x_1,-x_2)\quad\Leftrightarrow\quad \phi\mapsto\phi+\pi,\\
    &s_2:\,\varphi_2\mapsto\varphi_2+\pi,\quad
    (x_1,x_3)\mapsto(-x_1,-x_3)\quad\Leftrightarrow\quad \phi\mapsto-\phi+\pi,\,\tau\mapsto-\tau.
\end{aligned}
\end{equation}

The appearance of a hyperbolic space in the limiting UV geometry might seem unexpected. The explanation is as follows. There exists an element of the complexified gauge group $V\in \SO(3,\mathbb{C})$, explicitly given by
\begin{equation}
    V=e^{\left(i\rho+\pi/2\right)t_{2,3}}
    e^{\left(\phi+i\tau+\pi/2\right)t_{1,2}},
\end{equation}
where $t_{i,j}$ are the standard generators of $\mathfrak{so}(3)$, such that the product $e^{-\varphi_1 h_1-\varphi_2 h_2}gV$ is upper-triangular with positive diagonal entries
\begin{equation}
    \Omega=
    e^{-\varphi_1 h_1-\varphi_2 h_2}gV=
    \begin{pmatrix}
        \omega_1&\star&\star\\
        0&\omega_2&\star\\
        0&0&\omega_3
    \end{pmatrix},\qquad \omega_k\in\mathbb{R}_+
\end{equation}
The matrix $\Omega$ may be treated as an upper-triangular matrix factor of the Cholesky decomposition of the Hermitian positive-definite matrix $H$
\begin{equation}
    H=\Omega^\dagger\Omega.
\end{equation}
On the other hand, by construction we have $H = V^\dagger V\in \SO(3,\mathbb{C})$. That is, $H$ is a \textit{positive element} (i.e. Hermitian positive-definite) of $\SO(3, \mathbb{C})$, and thus it can be interpreted as an element of the coset space $\SO(3,\mathbb{C})/\SO(3,\mathbb{R})$ (see Section \ref{UV-theory} for more details), which is known to coincide with hyperbolic space $\mathbb{H}^3_+$
\begin{equation}
    H\in \SO(3,\mathbb{C})/\SO(3,\mathbb{R}) \simeq
    \mathbb{H}^3_+.
\end{equation}
\section{UV limit of \texorpdfstring{$\SU(N)/\SO(N)$}{SU(N)/SO(N)} YB-deformed sigma model}\label{UV-theory}

\paragraph{Parametrization of the coset element.} Motivated by the $\SU(3)/\SO(3)$ example, we conjecture the following ansatz for $\SU(N)/\SO(N)$ coset representative $g$
\begin{equation}\label{g-coset-parametrization}
    g=e^\Lambda\,\Omega\, V^{-1},\qquad\text{where}\qquad e^\Lambda \text{ is unitary diagonal},\quad V\in \mathrm{SO}(N,\mathbb C),
\end{equation}
and
\begin{equation}
    \Omega \text{ is upper-triangular with strictly positive real entries on the diagonal}.
\end{equation}
We emphasize that $V$ should be regarded as an element of $\SO(N,\mathbb{C})$ rather than $\SO(N,\mathbb{R})$. 

In this setup, such a decomposition of $g\in\SU(N)$ is just a complex analogue of the QR decomposition. Recall that the isotropy group $\SO(N)$ acts on $g$ from the right. Then the upper-triangular matrix $\Omega$, in fact, does not depend on the choice of a representative $g$ of an orbit $[g]\in\SU(N)/\SO(N)$. Indeed, a right ``gauge'' action $g\mapsto gO$, $O\in \SO(N, \mathbb{R})$ can always be compensated by the appropriate change of $V\in\SO(N,\mathbb{C})$, $V\mapsto O^{-1}V$, which does not spoil the orthogonality. More explicitly, 
\begin{equation}\label{g-coset-parametrization-3}
    gO=e^\Lambda\,\Omega \underbrace{(O^{-1}V)^{-1}}_{\in\,\SO(N,\mathbb{C})}.
\end{equation}

It is clear that a matrix $g\in\SU(N)$ in general position can always be decomposed as \eqref{g-coset-parametrization}. However, if $g$ belongs to a certain special subset of $\SU(N)$, such a decomposition is impossible. More precisely, by ``$g$ is in general position'' we actually mean that $[g]\in\SU(N)/\SO(N)$ lies in the intersection of $\SU(N)/\SO(N)$ with a so-called largest open Bruhat-type stratum of $\mathrm{SL}(N,\mathbb C)/\SO(N,\mathbb{C})$. We will denote this intersection by $\Sigma_e$ and call it ``the main stratum''. $\Sigma_e$ is known to be open and dense in $\SU(N)/\SO(N)$. In Section \ref{geometry} we discuss this in more detail.

\paragraph{Positive elements in $\SO(N, \mathbb{C})$.} For the next step, we define
\begin{equation}
    H=\Omega^\dagger\Omega.
\end{equation}
This matrix is Hermitian and positive-definite by construction. Moreover,
\begin{equation}
    H=\Omega^\dagger\Omega = \left(V^\dagger g^\dagger e^{\Lambda}\right)e^{-\Lambda}gV= V^\dagger V.
\end{equation}
$V$ is an element of $\SO(N, \mathbb{C})$; $V^\dagger = (V^{-1})^*$ is also an element of $\SO(N, \mathbb{C})$ (the orthogonal bilinear form is real, so complex conjugation preserves orthogonality). Hence, $H \in \SO(N, \mathbb{C})$ as well.

Now, there is a bijection between the set of positive elements inside $\SO(N, \mathbb{C})$ and the quotient $\SO(N,\mathbb{C})/\SO(N,\mathbb{R})$:
\begin{equation}
    \frac{\SO(N,\mathbb{C})}{\SO(N,\mathbb{R})} \quad \xleftrightarrow{\quad 1:1\quad} \quad \Big\{\text{positive elements in } \SO(N,\mathbb{C})\Big\}.
\end{equation}
The identification is given by the (unique) polar decomposition. For $V\in\SO(N, \mathbb{C})$ we write
\begin{equation}
    V = QP,\qquad Q \text{ unitary}, \quad P \text{ positive }.
\end{equation}
The polar factors are (uniquely) determined by $P = \sqrt{V^\dagger V}$, $Q = VP^{-1}$. Now, since $Q$ is both unitary and orthogonal, $Q$, in fact, lies in $\mathrm{U}(N)\cap\SO(N,\mathbb{C}) = \SO(N, \mathbb{R})$. In other words, when applied to a complex orthogonal matrix $V$, the polar decomposition writes $V$ uniquely as the product of a positive element $P\in\SO(N,\mathbb{C})^\textsf{pos.}$ and a real element $Q\in\SO(N,\mathbb{R})$. Then, the map $\SO(N, \mathbb{C})\longrightarrow\SO(N,\mathbb{C})^\textsf{pos.}$ given by
\begin{equation}
    V\quad \longmapsto \quad \sqrt{V^\dagger V}
\end{equation}
is clearly $\SO(N,\mathbb{R})$-invariant\footnote{Indeed, for any real orthogonal $O\in\SO(N,\mathbb{R})$ we have
\begin{equation}
    OV\quad\longmapsto\quad \sqrt{(OV)^\dagger(OV)} = \sqrt{V^\dagger (O^\mathrm{t}O)V} = \sqrt{V^\dagger V}
\end{equation}.}, and thus descends onto the quotient $\SO(N, \mathbb{C})/\SO(N, \mathbb{R})$. The uniqueness of the polar decomposition ensures bijectivity.

\paragraph{The UV fixed point geometry.} Recall that we have denoted the subset of $\SU(N)$ on which our initial decomposition
\begin{equation}
    g = e^\Lambda \Omega V^{-1}
\end{equation}
is well-defined, by $\Sigma_e\subset\SU(N)$. Now, given $g\in\Sigma_e$, this decomposition is not unique: if $V$ satisfies it, then so does $VD$ for any matrix $D$ that is both orthogonal ($\SO(N, \mathbb{C})$) and upper-triangular with real diagonal. There are, in general, multiple such $D$'s:
\begin{equation}
    D=\operatorname{diag}(\varepsilon_1,\ldots,\varepsilon_N),
    \qquad
    \varepsilon_i=\pm1,\qquad
    \prod_{i=1}^N\varepsilon_i=1.
\end{equation}
Such matrices $D$ form a discrete abelian group $\Gamma\simeq (\mathbb Z_2)^{N-1}$. In other words, $[g]\in\SU(N)/\SO(N)$ determines a pair $(e^\Lambda, H)$ only up to the action of $\Gamma$:\footnote{
Indeed, let us count the dimensions:
\begin{equation}
    \begin{aligned}
        \dim\left(\Sigma_e\right) &= N^2-1-\frac{N(N-1)}{2} = \frac{(N-1)(N+2)}{2};\\
        \dim\left(\mathrm{U}(1)^{N-1}\times \frac{\SO(N,\mathbb{C})}{\SO(N,\mathbb{R})}\right) &= N-1 + \frac{N(N-1)}{2} = \frac{(N-1)(N+2)}{2}.
    \end{aligned}
\end{equation}
The numbers of real parameters match.
}
\begin{equation}
  (e^\Lambda,H) \quad\overset{D}{\longmapsto}\quad (e^\Lambda D, DHD).
\end{equation}
Consider a map
\begin{equation}
    \Psi:\left(\mathrm{U}(1)^{N-1} \times \frac{\SO(N,\mathbb{C})}{\SO(N,\mathbb{R})}\right)\longrightarrow \frac{\SU(N)}{\SO(N)}
\end{equation}
sending
\begin{equation}
    (e^\Lambda, H) \longmapsto \big[e^\Lambda\Omega_H H^{-1/2}\big],
\end{equation}
where $\Omega_H$ is the (unique) Cholesky factor of $H$, an upper-triangular matrix with positive diagonal. The fibers of this $\Psi$ are precisely the $\Gamma$-orbits inside the space of $(e^\Lambda, H)$, so, $\Psi$ is a $2^{N-1}$-sheeted covering of $\Sigma_e$ (the main stratum, i.e. an open subset of $\SU(N)$ where $\Psi$ lands).

Denote by
\begin{equation}
    h_k=i(E_{kk}-E_{k+1,k+1}),
\qquad k=1,\ldots,N-1,
\end{equation}
the skew-Hermitian Cartan generators of $\mathfrak{su}(N)$ (essentially, these are traceless purely imaginary diagonal matrices). Then we can write
\begin{equation}
e^\Lambda=
\exp\left(\sum_{k=1}^{N-1}\alpha_k h_k\right),\quad \alpha_k\in[0,2\pi).
\end{equation}
Thus, since on the open stratum $\Sigma_e$ the element $g$ determines all ``phases'' $\alpha_k\in[0,2\pi)$, $k=1,\dots,N-1$ and the coset element $H\in \mathrm{SO}(N,\mathbb{C})/\mathrm{SO}(N,\mathbb{R})$ \textit{up to the action of} $\Gamma$, we have the identification
\begin{equation}\label{discrete-group-action}
(\alpha_k+\pi,H)
\sim
(\alpha_k,D_kHD_k),
\end{equation}
where matrices $D_k=e^{\pi h_k}$ are the generators of $\Gamma\simeq\mathbb{Z}_2^{N-1}$, the diagonal matrices with two $-1$'s on the diagonal, while the remaining diagonal entries are $1$.

\bigskip

For $N=2$, we have only one such matrix, $D_1 = \mathrm{diag}(-1, -1)$. Clearly, its action stabilizes $H$ (since $D_1$ is central), $D_1HD_1=H$. In that special case, the identification actually only halves the period of $\alpha_1$ from $2\pi$ to $\pi$. For $N\geq3$, the matrices $D_k$ are generally non-central and act non-trivially on $H$ as well. In any case, we have a quotient by a finite group:
\begin{equation}\label{finite-quotient}
    \Sigma_e \quad\simeq\quad \left(\mathrm{U}(1)^{N-1}\times \frac{\mathrm{SO}(N,\mathbb C)}{\mathrm{SO}(N,\mathbb R)}\right)\Big/(\mathbb Z_2)^{N-1}.
\end{equation}

\paragraph{The UV metric and $B$-field. } The parameterization \eqref{g-coset-parametrization} turns out to be very useful. The element $V$ in \eqref{g-coset-parametrization} belongs to the complexification of the gauge group $\SO(N, \mathbb{C})$. Thus in the action \eqref{Coset-action-deformed} one can ignore the $V^{-1}$ on the right and write
\begin{equation}
    g = e^{\Lambda}\Omega.
\end{equation}
Second, we note that from \eqref{R_g-definition} we have for the Drinfeld--Jimbo R-matrix \eqref{DJ-matrix}
\begin{equation}
    \mathcal{R}_{e^{\Lambda}\Omega}=\mathcal{R}_{\Omega},
\end{equation}
since $\mathcal{R}$ commutes with the adjoint action of the Cartan subalgebra. 

We have found the following relations:
\begin{equation}\label{relation-1}
    \left(1+i\mathcal{R}_{\Omega}\right)
\left(t\right)=\Omega^{-1}\,t\,\Omega\quad\text{for }\quad t\in\textrm{Cartan}\big(\mathfrak{su}(N)\big)
\end{equation}
and
\begin{equation}\label{relation-2}
    \left(1+i\mathcal{R}_{\Omega}\right)\mathcal{R}
    \left(H^{-1}dH\right)=2i\,\Omega^{-1}d\Omega-
    2i\,H^{-1}dH.
\end{equation}
We provide proofs of these formulas in Appendix \ref{relations-proof}.

From \eqref{relation-1} and \eqref{relation-2}, it immediately follows that the current $j=g^{-1}dg$ satisfies
\begin{equation}\label{main-equation}
    J=\frac{1}{1+i\mathcal{R}_{\Omega}\circ\mathrm{P}_{1}}\,
   j^{(1)}=
   d\Lambda+\frac{1}{2i}
   \mathcal{R}\left(H^{-1}dH\right).
\end{equation}
With the help of \eqref{main-equation}, it is now straightforward to calculate the metric and the $B$-field in the limit $\kappa\rightarrow 1$.

We start with the metric. For $\kappa=1$, it is given by
\begin{equation}
     ds^2|_{\kappa=1}=-\frac{1}{2}\textrm{Tr}\left(J\,J\right).
\end{equation}
Substituting \eqref{main-equation}, the metric for $\kappa=1$ reads
\begin{equation}
    ds^2|_{\kappa=1}=-\frac{1}{2}\textrm{Tr}\left(d\Lambda^2\right)+\frac{1}{8}\textrm{Tr}\left(H^{-1}dH\,H^{-1}dH\right).
\end{equation}

Next, we calculate the $B$-field
\begin{equation}
    B=-\frac{i}{2} \textrm{Tr}\left(J\wedge \mathrm{P}_{1}\circ\mathcal{R}_{g}\circ\mathrm{P}_{1}\left(J\right)\right).
\end{equation}
Applying \eqref{relation-1}, \eqref{relation-2} and \eqref{main-equation}, one gets the following formula
\begin{equation}
    B=-\frac{i}{2} \textrm{Tr}\left(\left(d\Lambda+\frac{1}{2i}
   \mathcal{R}\left(H^{-1}dH\right)\right)\wedge \left(-i\, \mathrm{P}_{1}\left(\Omega^{-1}d\Lambda\Omega-d\Lambda\right)-\frac{1}{2}\left(i\mathrm{P}_{1}\left(\Omega^{-1}d\Omega\right)-\mathcal{R}\left(H^{-1}dH\right)\right)\right)\right).
\end{equation}
The terms
\begin{equation}
    \textrm{Tr}\left(d\Lambda \wedge \mathrm{P}_{1}\left(\Omega^{-1}d\Lambda\Omega-d\Lambda\right)\right) \qquad\text{and}\qquad \textrm{Tr}\left(d\Lambda \wedge \mathcal{R}\left(H^{-1}dH\right)\right)
\end{equation}
are identically equal to zero. Additionally, the following forms are exact:
\begin{equation}
    \textrm{Tr}\left(H^{-1}dH\wedge H^{-1}dH\right)=-d\left(\textrm{Tr}\left(H^{-1}dH\right)\right),\quad\textrm{Tr}\left(d\Lambda\wedge  \mathrm{P}_{1}\left(\Omega^{-1}d\Omega\right)\right)=d\left(\log{(\textrm{Diag}(\Omega))}\,d\Lambda\right)
\end{equation}
and 
\begin{equation}
    \textrm{Tr}\left(\mathcal{R}\left(H^{-1}dH\right)\wedge\mathrm{P}_{1}\left(\Omega^{-1}d\Lambda\Omega-d\Lambda\right)\right)=-d\left(\left(\log{(\textrm{Diag}(\Omega))}+\log{(\textrm{Diag}(\Omega^{\dagger}))}\right)\,d\Lambda\right).
\end{equation}

Thus, up to an exact form $d\omega$, 
\begin{equation}
    B=\frac{1}{8} \textrm{Tr}\left(
   H^{-1}dH\wedge \Omega^{-1}d\Omega\right)+d\omega.
\end{equation}
Hence,
\begin{equation}
    \mathcal{H}=dB=-\frac{1}{8} \textrm{Tr}\left(
   H^{-1}dH\wedge H^{-1}dH\wedge\Omega^{-1}d\Omega\right)+\frac{1}{8} \textrm{Tr}\left(
   H^{-1}dH\wedge \Omega^{-1}d\Omega\wedge\Omega^{-1}d\Omega\right).
\end{equation}
Additionally, the following property holds for the upper-triangular matrix $\Omega$:
\begin{equation} \label{triple-omega}
    \textrm{Tr}\left(\Omega^{-1}d\Omega \wedge \Omega^{-1}d\Omega \wedge \Omega^{-1}d\Omega\right)=0.
\end{equation}
Using \eqref{triple-omega}, the expression for $\mathcal{H}$ can be put in a more convenient form:
\begin{equation}
    \mathcal{H}=-\frac{1}{24}\textrm{Tr}\left(
   H^{-1}dH\wedge H^{-1}dH\wedge H^{-1}dH\right).
\end{equation}

\paragraph{The vicinity of the UV fixed point.} Now we consider the action on $\SU(N)/\SO(N)$ near the UV point $\kappa=1$:
\begin{equation}
    \mathcal{S}=-\frac{1}{2\hbar}\int\textrm{Tr}\left(j_{+}^{(1)}\,\frac{1}{1+i\mathcal{R}_{g}\circ\mathrm{P}_{1}}\,
   j_{-}^{(1)}\right) d^{2}x-
   \frac{\kappa-1}{2\hbar}\int\textrm{Tr}\left( j_{+}^{(1)}\,\frac{1}{\left(1+i\mathcal{R}_{g}\circ\mathrm{P}_{1}\right)^2}\,
   j_{-}^{(1)}\right) d^{2}x+O\left((\kappa-1)^2\right).
\end{equation}
By \eqref{main-equation}, in order to compute the first-order term, it is enough to calculate the expression
\begin{equation}
    j^{(1)} \frac{1}{1+i\mathcal{R}_{\Omega}\circ\mathrm{P}_{1}}.
\end{equation}
To this end, we notice that, from the trace pairing, one has the following equation:
\begin{equation}\label{main-equation-transposed}
    j_{\Omega}^{(1)} \frac{1}{1-i\mathcal{R}_{\Omega}\circ\mathrm{P}_{1}}\,=
    d\Lambda+\frac{1}{2i}
   \mathcal{R}\left(H^{-1}dH\right).
\end{equation}
Here we explicitly indicate the dependence of $j_{\Omega}$ on the upper-triangular $\Omega$
\begin{equation}
j_{\Omega}^{(1)} = \left(e^{\Lambda}\Omega\right)^{-1} d \left( e^{\Lambda}\Omega\right).
\end{equation}

Now, consider instead the lower-triangular matrix $\bar{\Omega}$ and parameterize the group element $g$ as $g=e^{\Lambda}\bar{\Omega}$. It is easy to check that \eqref{main-equation-transposed} becomes 
\begin{equation}
    j_{\bar{\Omega}}^{(1)} \frac{1}{1+i\mathcal{R}_{\bar{\Omega}}\circ\mathrm{P}_{1}}\,=
   d\Lambda-\frac{1}{2i}
   \mathcal{R}\left(\bar{H}^{-1}d\bar{H}\right),
\end{equation}
where now 
\begin{equation}
j_{\bar{\Omega}}^{(1)} = \left(e^{\Lambda}\bar{\Omega}\right)^{-1} d \left( e^{\Lambda}\bar{\Omega}\right),\quad \bar{H}=\bar{\Omega}^{\dagger}\bar{\Omega}.
\end{equation}
We can take $\bar{\Omega}=\Omega H^{-1}={\left(\Omega^{+}\right)}^{-1}$ and, consequently, $\bar{H}=H^{-1}$. The matrices $\Omega$ and $\bar{\Omega}$ are related by the action of the $\SO(N,\mathbb{C})$ element $H$. Hence,
\begin{equation}
    j_{\Omega}^{(1)} \frac{1}{1+i\mathcal{R}_{\Omega}\circ\mathrm{P}_{1}}\,=
   H^{-1}\left(j_{\bar{\Omega}}^{(1)} \frac{1}{1+i\mathcal{R}_{\bar{\Omega}}\circ\mathrm{P}_{1}}\right)H.
\end{equation}

Finally, we arrive at the following formula for the metric (divided by $4$)
\begin{equation}\label{metric-first-correction}
ds^2=ds^2|_{\kappa=1}-\frac{(\kappa-1)}{2}\,\textrm{Tr}\left(H^{-1}\Bigl(d\Lambda - \frac{1}{2i} \mathcal{R}\left(HdH^{-1}\right)\Bigr)H \Bigl(d\Lambda + \frac{1}{2i}
\mathcal{R}\left(H^{-1}dH\right)\Bigr)\right)+O\left((\kappa-1)^2\right),
\end{equation}
and for the B-field
\begin{equation}\label{B-field-first-correction}
B=B|_{\kappa=1}-\frac{(\kappa-1)}{2}\,\textrm{Tr}\left(H^{-1}\Bigl(d\Lambda - \frac{1}{2i} \mathcal{R}\left(HdH^{-1}\right)\Bigr)H \wedge\Bigl(d\Lambda + \frac{1}{2i}
\mathcal{R}\left(H^{-1}dH\right)\Bigr)\right)+O\left((\kappa-1)^2\right).
\end{equation}
\section{First-order term and screening operators}\label{screenings}
The direct quantization of the Yang--Baxter-deformed sigma model is a subtle problem. One has to control the simultaneous renormalization of the metric and the $B$-field, possible field redefinitions, and higher-loop corrections to the background. It is therefore natural to formulate the quantum theory in terms of conformal perturbation theory around its ultraviolet fixed point. This approach has been used in \cite{Fateev:1996ea,Fateev:2017mug,Fateev:2018yos,Litvinov:2018bou,Fateev:2019xuq,Litvinov:2019rlv,Alfimov:2020jpy}. The first step is to identify the conformal field theory described by the leading metric and $B$-field. The sub-leading terms in the ultraviolet expansion should then be interpreted as the semiclassical limits of relevant operators perturbing this CFT. Quantum corrections to the perturbing operator can be constrained by requiring the conservation of local integrals of motion.

The leading ultraviolet metric and $B$-field in \eqref{metric-first-correction}--\eqref{B-field-first-correction} are not merely reminiscent of a WZW background; together they reproduce the two geometric terms entering the WZW functional \cite{Witten:1983ar}. Indeed, introducing the $\SO(N,\mathbb C)/\SO(N,\mathbb R)$-valued Hermitian field $H$ and the Maurer--Cartan current
\begin{equation}
J=H^{-1}dH,
\end{equation}
the non-abelian part of the limiting metric and $B$-field \eqref{metric-first-correction}--\eqref{B-field-first-correction} takes the form
\begin{equation}
ds^2
=\frac{1}{8}\operatorname{Tr}(J^2),
\qquad
\mathcal \mathcal{H}=dB
=-\frac{1}{24}\operatorname{Tr}(J\wedge J\wedge J).
\label{WZW-background}
\end{equation}
The first expression is the invariant metric on the non-compact symmetric space $\SO(N,\mathbb C)/\SO(N,\mathbb R)$,
whereas the second is its canonical Wess--Zumino three-form. Most importantly, the relative normalization of the metric and the $B$-field in \eqref{WZW-background} is precisely the one required by the WZW action. The corresponding quantum Euclidean action can therefore be written as
\begin{equation}\label{noncompact-WZW-action}
S_{\rm WZW}[H]
=
\frac{k}{8\pi}
\int
\operatorname{Tr}
\bigl(
H^{-1}\partial H
H^{-1}\bar\partial H
\bigr)d^2z
-
\frac{i k}{24\pi}
\int
\operatorname{Tr}
\bigl(H^{-1}dH\bigr)^{\wedge 3}.
\end{equation}

We briefly comment on the fact that the $B$-field in \eqref{noncompact-WZW-action} is imaginary. This is a characteristic feature of the Euclidean non-compact real form and follows directly from the reality condition $H^\dagger=H$. In this case, the three-form is purely imaginary
\begin{equation}
\operatorname{Tr}(J\wedge J\wedge J)^*
=-\operatorname{Tr}(J\wedge J\wedge J),
\end{equation}
and thus the corresponding factor of $i$ makes the action \eqref{noncompact-WZW-action} real.

Thus, once the overall coefficient in \eqref{noncompact-WZW-action} is identified with the level $k$, quantization of the leading non-abelian background gives the non-compact WZW model associated with
$\SO(N,\mathbb C)/\SO(N,\mathbb R)$ \cite{Gawedzki:1991yu}. The Cartan part of the UV metric \eqref{metric-first-correction} is decoupled from $H$ and has a vanishing $B$-field (up to exact form)
\begin{equation}
ds^2_{\rm Cartan}=-\frac{1}{2}\operatorname{Tr}(d\Lambda)^2 .
\end{equation}
It therefore quantizes to $N-1$ compact free bosons. Locally, the ultraviolet conformal field theory is consequently
\begin{equation}\label{local-UV-CFT}
\frac{\SO(N,\mathbb C)}{\SO(N,\mathbb R)}
\ \text{WZW}
\ \otimes \ 
\mathrm{U}(1)^{N-1}.
\end{equation}
Globally, one should additionally take into account the residual finite identifications \eqref{finite-quotient}. This identification does not change the local WZW geometry, the local current algebra, or the central charge. Generally speaking, completing the quantum identification requires determining the level $k$ in terms of the parameter $\hbar$ of the initial action \eqref{Coset-action-deformed} and the compactification lattice of the abelian bosons.

The first-order correction terms in \eqref{metric-first-correction} and \eqref{B-field-first-correction} correspond to the  perturbation of the model \eqref{local-UV-CFT} by the following operator
\begin{equation}
\lambda \int \,\textrm{Tr}\left(H^{-1}\bar{A}HA\right)d^2x,
\end{equation}
where
\begin{equation}
    A=\partial\Lambda+\frac{1}{2i}\mathcal{R}
   \left(H^{-1}\partial H\right),\quad
   \bar{A}=\bar{\partial}\Lambda-\frac{1}{2i}\mathcal{R}
   \left(H\bar{\partial}H^{-1}\right).
\end{equation}
At the quantum level, one identifies this perturbation with the symmetric-traceless rank-two WZW primary
\begin{equation}\label{our-perturbation}
    \int \eta_m\bar\eta_{\bar m}\,\Phi^{(2)}_{m,\bar m}\ d^2x,
\end{equation}
with
\begin{equation}
    \eta=\partial\Lambda+\frac{\gamma}{2i}\,\mathcal{R}
   \left(J\right),\quad
   \bar{\eta}=\bar{\partial}\Lambda-\frac{\gamma}{2i}\,\mathcal{R}
   \left(\bar{J}\right),
\end{equation}
where we use the affine currents
\begin{equation}
J=k\,H^{-1}\partial H,\quad
\bar{J}=k\,H\partial H^{-1},
\end{equation}
and CFT ordering of fields is assumed in \eqref{our-perturbation}. The parameter $\gamma$ in \eqref{our-perturbation} is a quantization parameter such that
\begin{equation}
    k\gamma^2\to 1\quad\text{at}\quad k\to\infty.
\end{equation}

The perturbation \eqref{our-perturbation} is reminiscent of the integrable perturbations of the $\SO(3)$ WZW model studied by Brazhnikov \cite{Brazhnikov:1996fa} and generalized by Castro Alvaredo and Miramontes for maximal-rank symmetric spaces \cite{CastroAlvaredo:2000kq} (see Section \ref{generalizations} for more details). For the symmetric-space decomposition of the compact simple Lie algebra for $\mathfrak{g}=\mathfrak{g}^{(0)}\oplus \mathfrak{g}^{(1)}$, the corresponding action takes the form
\begin{equation}\label{Miramontes-perturbation}
    S=S_{\mathrm{WZW}}[h]
    +\mu
    \int d^2x\,
    \left\langle
       \Lambda_+,
       h^\dagger\Lambda_-h
    \right\rangle,
\end{equation}
where $h\in H$ and $\Lambda_\pm$ are regular Cartan elements of $\mathfrak{g}$ (for maximal rank symmetric spaces of $\mathfrak{g}^{(1)}$). Quantum integrability was investigated within the conformal perturbation approach by showing that there are local conserved currents that survive the perturbation. Each conserved current defines the integral of motion 
\begin{equation}\label{local-IM}
    I_s=\int G_{s+1}(z)dz,
\end{equation}
which commutes \cite{Zamolodchikov:1989zs} with the chiral part of perturbation \eqref{Miramontes-perturbation}. In particular, in addition to the Sugawara density \cite{Brazhnikov:1996fa}
\begin{equation}
    G_2^{(1)}=\mathrm{Tr}(JJ),
\end{equation}
there is an additional density of spin $2$
\begin{equation}\label{Brazhnikov-density}
    G_2^{(2)}=\mathrm{Tr}(\Lambda_-JJ).
\end{equation}
The models \eqref{Miramontes-perturbation} were called symmetric space sine-Gordon models (SSSG).

The chiral part $\Phi^{(2)}(z)$ of the operator $\Phi^{(2)}(z,\bar{z})$ is a WZW primary in the symmetric traceless square $\operatorname{Sym}^{2}_{0}(\mathbb C^{N})$ of the $\SO(N)$ vector representation whose highest weight is twice that of the vector representation. The components of  $\Phi^{(2)}(z)$ are labeled by symmetric traceless matrices
\begin{equation}\label{Phi2-components}
    \Phi^{(2)}(z)=\sum s_{ij}\Phi^{(2)}_{s_{ij}}(z),\quad s_{ij}=E_{ij}+E_{ji},\quad
    s_{ii}=E_{ii}-\frac{1}{N}\sum_{j=1}^NE_{jj},
    \quad
    \sum_{i=1}^Ns_{ii}=0.
\end{equation}
Let
\begin{equation}
    t_{ij}=E_{ij}-E_{ji}\quad\text{for}\quad i<j,
\end{equation}
be the $\mathfrak{so}(N)$ generators, so that $J(z)=\sum_{i<j}J_{ij}(z)t_{ij}$. Currents $J_{ij}(z)$ obey the OPE
\begin{equation}
    J_{ij}(z)J_{mn}(w)=\frac{-k\,\delta_{im}\delta_{jn}}{(z-w)^2}+
    \frac{\delta_{jm}J_{in}(w)-\delta_{im}J_{jn}(w)-\delta_{jn}J_{im}(w)+\delta_{in}J_{jm}(w)}
    {z-w}+\dots
\end{equation}
Then the OPE of \eqref{Phi2-components} with WZW currents $J_{ij}(z)$ has the form
\begin{equation}
    J_{ij}(z)\Phi^{(2)}(w)=
    -\frac{[t_{ij},\Phi^{(2)}(w)]}{z-w}+\dots
\end{equation}
Also, we have
\begin{equation}
\mathcal{R}(J)=i\sum_{i<j}J_{ij}s_{ij}.
\end{equation}

We define the screening charge
\begin{equation}\label{screening-proposal}
    S=\oint\Tr\left(\Bigl(\partial\Lambda+\frac{\gamma}{2i}\mathcal{R}(J)\Bigr)\Phi^{(2)}\right)dz
\end{equation}
In order to rewrite \eqref{screening-proposal} in components, we define an orthogonal basis
\begin{equation}
    d_i=\sqrt{\frac{2}{i(i+1)}}\operatorname{diag}(1^i,-i,0^{N-i-1}),\quad
    \Tr(d_id_j)=2\delta_{ij},
\end{equation}
and the free currents
\begin{equation}
    \partial\Lambda=\frac{1}{2}\sum_{i=1}^{N-1}\partial\varphi_id_i,
    \quad
    \partial\varphi_i(z)\partial\varphi_j(w)=-\frac{\delta_{ij}}{(z-w)^2}+\dots
\end{equation}
Then the screening \eqref{screening-proposal} becomes
\begin{equation}\label{screening-proposal-2}
    S=
    \oint \left(
    \sum_{i=1}^{N-1}\partial\varphi_i(z)\Phi^{(2)}_{d_{i}}(z)
    +\gamma
    \sum_{i<j}J_{ij}(z)\Phi_{s_{ij}}^{(2)}(z)\right)dz.
\end{equation}
Normal ordering of operator products is understood. A local integral of motion \eqref{local-IM} is said to commute with the screening charge if the OPE residue with its density is a total derivative
\begin{equation}
    \oint_{\mathcal{C}_w} \mathcal{V}(z)G_3(w)dz=\partial R(w),
\end{equation}
where $\mathcal{V}(z)$ is the integrand in \eqref{screening-proposal}.

We note that the integrand in \eqref{screening-proposal} is defined modulo total derivatives that are controlled by the KZ equation
\begin{equation}\label{KZ-equation}
    (k+N-2)\,\partial\Phi^{(2)}(z)=
     \sum_{i<j}J_{ij}\bigl[t_{ij,}\Phi^{(2)}(z)\bigr].
\end{equation}
Taking the trace of \eqref{KZ-equation} with
\begin{equation}
    C=\operatorname{diag}(c_1,\ldots,c_N),
    \quad \sum_i c_i=0,
\end{equation}
and using $[C,t_{ij}]=(c_i-c_j)s_{ij}$, we get
\begin{equation}\label{total-derivative-redundancy}
    \sum_{i<j}(c_i-c_j)J_{ij}\Phi_{s_{ij}}^{(2)}(z)\simeq0.
\end{equation}

For $N=2$ the redundancy \eqref{total-derivative-redundancy} eliminates completely the $\mathcal{R}(J)$ part of the screening, and hence the quantization parameter $\gamma$ remains unfixed. For $N=3$ only one component survives
\begin{equation}
    P_{\perp}=\frac{1}{3}\left(
J_{12}(z)\Phi_{s_{12}}^{(2)}(z)
-J_{13}(z)\Phi_{s_{13}}^{(2)}(z)
+J_{23}(z)\Phi_{s_{23}}^{(2)}(z)
\right).
\end{equation}
Then the screening charge \eqref{screening-proposal} becomes
\begin{equation}\label{screening-proposal-3}
\mathcal{S}_{N=3}=
\oint\left(\partial\varphi_1(z)\Phi_{d_1}^{(2)}(z)+
\partial\varphi_2(z)\Phi_{d_2}^{(2)}(z)
+\gamma P_{\perp}(z)\right)dz.
\end{equation}

The case $N=2$ is exceptional: the first non-trivial integral of motion has spin $3$ \cite{Litvinov:2018bou}. For $N\geq3$, we conjecture that there exists a non-trivial integral of spin $2$, with the corresponding density of spin $3$.  Its form is restricted by the discrete symmetry group \eqref{discrete-group-action}. For $N=3$, the discrete group $\mathbb Z_2\times\mathbb Z_2$ acts by
\begin{equation}
D=\operatorname{diag}(\epsilon_1,\epsilon_2,\epsilon_3),\qquad
\epsilon_i=\pm1,\quad \epsilon_1\epsilon_2\epsilon_3=1.
\end{equation}
It conjugates the $\SO(3)$ field $H$ and shifts the compact bosons $\varphi_k$. Consequently $\partial\varphi_k$ are unchanged, while
\begin{equation}
J_{ij}\longmapsto\epsilon_i\epsilon_j\,J_{ij}.
\end{equation}
Assuming that the IM current respects this symmetry, we find that among local densities of conformal weight three, the invariant ansatz, modulo total derivatives, is
\begin{equation}
G_3={}c_1\partial\varphi_1^3+c_2\partial\varphi_1^2\partial\varphi_2+c_3\partial\varphi_1\partial\varphi_2^2+c_4\partial\varphi_2^3
+\sum_{i<j}(a_{ij}\partial\varphi_1 J_{ij}^2+b_{ij}\partial\varphi_2 J_{ij}^2)
+c_5J_{12}J_{13}J_{23}+c_6\partial\varphi_1\partial^2\varphi_2 .
\end{equation}
Thus, we have $12$ undetermined coefficients (one coefficient fixes the overall normalization). In addition, we require the large zero-mode limit of compact bosons $\partial\varphi_k$ to coincide with Brazhnikov's density $G_2^{(2)}$ \eqref{Brazhnikov-density}. This requirement fixes
\begin{equation}
(a_{12},a_{13},a_{23},b_{12},b_{13},b_{23})
=c_7\left(0,1,-1,\frac{2}{\sqrt{3}},-\frac{1}{\sqrt{3}},-\frac{1}{\sqrt{3}},\right).
\end{equation}
This reduces the ansatz to seven coefficients $(c_1,c_2,c_3,c_4,c_5,c_6,c_7)$. They are to be determined from the condition
\begin{equation}\label{residue-condition}
    \oint_{\mathcal{C}_w} \mathcal{V}(z)G_3(w)dz=\partial R(w),
\end{equation}
where
\begin{equation}
    \mathcal{V}(z)=\partial\varphi_1(z)\Phi_{d_1}^{(2)}(z)+
\partial\varphi_2(z)\Phi_{d_2}^{(2)}(z)
+\gamma P_{\perp}(z).
\end{equation}

The residue condition \eqref{residue-condition} appears to have a non-trivial solution 
\begin{equation}
    (c_1,c_2,c_3,c_4,c_5,c_6,c_7)=
    \left(0,-\frac{1-2k}{\sqrt{3}},0,\frac{1-2k}{3\sqrt{3}},2\gamma,\frac{2\gamma(1-2k)}{\sqrt{3}},1\right)
    \quad\text{and}\quad
    \gamma=\pm\frac{1}{\sqrt{k-4}}.
\end{equation}
It is remarkable that after an overall rescaling by $1/2$, and modulo total derivatives, the density can be written in the following trace form
\begin{equation}\label{IM-invariant-3}
    G_3=\frac{2(2k-1)}{3}\Tr\left(\partial\Lambda^3\right)
    -\Tr\left(\partial\Lambda\mathcal{R}(J)^2\right)+
    \frac{i\gamma}{6}\Tr\left(\mathcal{R}(J)^3\right)-
    \frac{\gamma(2k-1)}{\sqrt{3}}
\partial\varphi_1\partial^2\varphi_2.
\end{equation}
\section{Geometry of \texorpdfstring{$\SU(N)/\SO(N)$}{SU(N)/SO(N)}}\label{geometry}

Let a point of $\SU(N)/\SO(N)$ be represented by $g\in \SU(N)$.  We ask whether there exists a complex orthogonal matrix $V\in \SO(N,\mathbb C)$ such that
\begin{equation}\label{orthogonal-triangularization}
  gV=b,
  \qquad b\in B_+,
\end{equation}
where $B_+\subset \mathrm{GL}(N,\mathbb C)$ is the upper-triangular Borel subgroup. This question is well defined on the compact coset $\SU(N)/\SO(N)$. Indeed, a change of representative $g\mapsto gO$, $O\in \SO(N)$ can be compensated by replacing
$V\mapsto O^{-1}V$. Thus, the whole $\SO(N)$-orbit containing $g$ is mapped to the same $b\in B_+$. It is important that $V$ is not a transformation by the compact gauge group $\SO(N)$, but belongs to its complexified version, $V\in\SO(N, \mathbb{C})$.

To interpret \eqref{orthogonal-triangularization}, set
\begin{equation}
  X_N^{\mathbb C}=\mathrm{GL}(N,\mathbb C)/\mathrm{O}(N,\mathbb C).
\end{equation}
The Cartan embedding identifies this complex symmetric space $X_N^\mathbb{C}$ with the space of symmetric invertible matrices inside $\GL(N, \mathbb{C})$:
\begin{equation}\label{Xc-Cartan-embedding}
  X_N^{\mathbb C}
  \simeq
  \bigl\{S\in \mathrm{GL}(N,\mathbb C):\ S^T=S\bigr\}.
\end{equation}
The compact symmetric space $X_N\subset X_N^\mathbb{C}$ is embedded as
\begin{equation}
  X_N=
  \bigl\{S\in \SU(N):\ S^T=S\bigr\} \subset X_N^\mathbb{C}.
  \label{eq:compact-cartan-realization}
\end{equation}
For the representative $g$, the corresponding Cartan-embedded matrix is $S=gg^T\in X_N$. Now, let $gV=b$, as in \eqref{orthogonal-triangularization}. Then the orthogonality of $V$ yields
\begin{equation}\label{cholesky-decomposition}
  S=gg^T=bb^T.
\end{equation}
Conversely, suppose that $S=gg^T=bb^T$ for some $b\in B_+$. Since $\det S=1$, we have $\det b=\pm1$. Set
\begin{equation}
D=\operatorname{diag}(\det b,1,\ldots,1).
\end{equation}
Replacing $b$ by $bD$ preserves both upper triangularity and the factorization $S=bb^T$ and ensures that $\det b=1$. With this choice,
$V=g^{-1}b$ satisfies 
\begin{equation}
VV^T=g^{-1}bb^T(g^T)^{-1}=1,
\qquad
\det V=1.
\end{equation}
Thus, $V\in \SO(N,\mathbb C)$ and $gV=b\in B_+$. We therefore obtain the equivalence
\begin{equation}\label{triang-open-orbit}
\exists\,V\in \SO(N,\mathbb C):\ gV\in B_+
\quad\Longleftrightarrow\quad
\exists\,b\in B_+:\ S=gg^T=bb^T.
\end{equation}

The r.h.s. of \eqref{triang-open-orbit} belongs to the orbit of $e\in\GL(N, \mathbb{C})$ under the action of the upper-triangular Borel subgroup $B_+\subset \GL(N,\mathbb{C})$, which acts on $X_N^{\mathbb{C}}$ by congruence
\begin{equation}\label{Xc-action}
S\longmapsto bSb^T.
\end{equation}
The orbits of this action, which are symmetric analogs of the Bruhat decomposition, were studied in \cite{RichardsonSpringer1990,BagnoCherniavsky2012}. The result is
\begin{equation}\label{symmetric-Bruhat}
  X_N^{\mathbb{C}}
  =\bigsqcup_{\substack{\sigma\in S_N \\ \sigma^2=1}}bP_{\sigma}b^T,\quad
  b\in B_+\subset\GL(N,\mathbb{C}),
\end{equation}
where  $P_{\sigma}\in\GL(N,\mathbb{C})$ is the matrix of the permutation $\sigma\in S_N$
\begin{equation}
  P_{\sigma} e_j=e_{\sigma(j)}.
\end{equation}
It is important that only involutions $\sigma^2=1$ appear in \eqref{symmetric-Bruhat}. Equivalently, for every invertible symmetric matrix $S\in \mathrm{GL}(N,\mathbb{C})$, there is a unique involution $\sigma\in S_N$ and some $b\in B_+$ such that
\begin{equation}\label{symmetric-Bruhat-2}
  bSb^{T}=P_{\sigma}.
\end{equation} 
The matrix $b\in B_+$ need not be unique. We note that the ordinary Bruhat decomposition is as follows
\begin{equation}\label{ordinary-Bruhat}
  \GL(N,\mathbb{C})=\bigsqcup_{w\in S_N}B_+P_wB_+,
\end{equation}
so that for any matrix $G\in \mathrm{GL}(N,\mathbb{C})$ there exists a unique permutation $\sigma\in S_N$ and two upper-triangular matrices $b_1,b_2\in B_+$ such that
\begin{equation}\label{ordinary-Bruhat-2}
b_1\,G\,b_2=P_{\sigma}.
\end{equation}
Here, the upper-triangular factors $b_1$ and $b_2$ are independent. In the symmetric version of this problem \eqref{symmetric-Bruhat-2}, the rightmost factor $b^T$ is lower-triangular and forced to be the transpose of the left factor $b\in B_+$. Consequently, the remaining permutation matrix $P_\sigma$ must itself be symmetric, so only involutions $\sigma^2=1$ survive. Since the decomposition \eqref{symmetric-Bruhat} is less familiar than the ordinary one, we provide a proof in Appendix \ref{symmetric-Bruhat-proof}.

For $1\le i,j\le N$, define the rank-control matrix \cite{BagnoCherniavsky2012}
\begin{equation}
  \rho_{ij}(S)
  =\textrm{rank}\left(S_{\{i,i+1,\ldots,N\},\{j,j+1,\ldots,N\}}\right).
\end{equation}
It can be easily shown that every $\rho_{ij}$ is invariant under upper-triangular congruence. So, the rank-control matrix can be used to distinguish the components $b$ and $b^T$ of \eqref{symmetric-Bruhat}:
\begin{equation}
    S\in B_+P_{\sigma}B_+^T\iff \rho_{ij}(S)=\rho_{ij}(P_{\sigma}).
\end{equation}

Now we intersect the symmetric Bruhat decomposition \eqref{symmetric-Bruhat} with $\SU(N)/\SO(N)$. It can be easily shown that the intersection of each stratum in \eqref{symmetric-Bruhat} with $\SU(N)/\SO(N)$ is non-empty. Let us define
\begin{equation}
    \Sigma_{\sigma}=\big\{S\in \SU(N):\quad
    S=S^T \quad \text{and} \quad
    \rho_{ij}(S)=\rho_{ij}(P_{\sigma})\quad\text{for all } i,j\big\}.
\end{equation}
Consequently,
\begin{equation}\label{SU(N)-SO(N)-strata}
    \SU(N)/\SO(N)=
    \bigsqcup_{\substack{\sigma \in S_N \\ \sigma^2=1}}\Sigma_{\sigma}.
\end{equation}
Note that $\Sigma_{\sigma}$ are not CW cells in the usual sense. Their topology can include torus directions and nontrivial finite identifications. Moreover, $\Sigma_{\sigma}$ need not be connected. 

Let us consider the lower-dimensional examples.
\paragraph{Case $N=2$.} In this case
\begin{equation}
        S=\begin{pmatrix}
           z&it\\
           it&\bar{z}
           \end{pmatrix},\quad
           t\in\mathbb{R},\quad |z|^2+t^2=1.
\end{equation}
Thus $\SU(2)/\SO(2)\simeq S^2$. There are two strata in \eqref {SU(N)-SO(N)-strata}, corresponding to
\begin{equation}
P_e=\begin{pmatrix}
   1&0\\
   0&1
\end{pmatrix}
\quad\text{and}\quad
P_{(12)}=\begin{pmatrix}
   0&1\\
   1&0
\end{pmatrix}.
\end{equation}
For the identity element, we have
\begin{equation}
\rho_e
=
\begin{pmatrix}
2&1\\
1&1
\end{pmatrix},
\end{equation}
and therefore $S$ with this control matrix has $z\neq0$. Thus
\begin{equation}
\Sigma_e
=\begin{pmatrix}
           z&it\\
           it&\bar{z}
 \end{pmatrix}\quad\text{with}\quad z\neq0.
\end{equation}
For non-trivial involution $\sigma=(12)$, one has
\begin{equation}
\rho_{(12)}
=
\begin{pmatrix}
2&1\\
1&0
\end{pmatrix},
\end{equation}
and therefore $z=0$ and $t=\pm1$.
Hence
\begin{equation}
\Sigma_{(12)}=\Sigma_{(12)}^{+}\sqcup\Sigma_{(12)}^{-},\quad\Sigma_{(12)}^{\pm}
=
\begin{pmatrix}
0&\pm i\\
\pm i&0
\end{pmatrix}.
\end{equation}
Thus
\begin{equation}
S^2
=
\Sigma_e
\sqcup
\Sigma_{12}^+
\sqcup
\Sigma_{12}^-.
\end{equation}
Notice that
\[
\Sigma_e
=
S^2\setminus\{p_+,p_-\}
\simeq
S^1\times\mathbb{R}.
\]
Thus the strata $\Sigma_{\sigma}$ are not CW cells. Moreover, $\Sigma_{(12)}$ has two components $\Sigma_{(12)}^{\pm}$.

\paragraph{Case $N=3$.} In this case
\begin{equation}
S=
\begin{pmatrix}
a&b&c\\
b&d&e\\
c&e&f
\end{pmatrix},
\end{equation}
where $(a,b,c,d,e,f)$ are constrained by the conditions $SS^{\dagger}=1$ and $\det S=1$. In particular,
\begin{equation}\label{S-in-N-3}
    \det\begin{pmatrix}
        d&e\\
        e&f
    \end{pmatrix}=\bar{a}\quad\text{and}\quad
    \det\begin{pmatrix}
        b&c\\
        e&f
    \end{pmatrix}=-\bar{b}.
\end{equation}
There are four strata corresponding to involutions $e$, $(12)$, $(23)$ and $(13)$ with the rank-control matrices
\begin{equation}
    \rho_{e}=\begin{pmatrix}
        3&2&1\\
        2&2&1\\
        1&1&1
    \end{pmatrix},\quad
    \rho_{(12)}=\begin{pmatrix}
        3&2&1\\
        2&1&1\\
        1&1&1
    \end{pmatrix},\quad
    \rho_{(23)}=\begin{pmatrix}
        3&2&1\\
        2&2&1\\
        1&1&0
    \end{pmatrix}\quad\text{and}\quad
    \rho_{(13)}=\begin{pmatrix}
        3&2&1\\
        2&1&0\\
        1&0&0
    \end{pmatrix}.
\end{equation}
Therefore, the four strata are 
\begin{equation}
    \Sigma_e=\{a\neq0,f\neq 0\},\quad
    \Sigma_{(12)}=\{a=0,b\neq0\},\quad
    \Sigma_{(23)}=\{f=0,e\neq0\}\quad\text{and}\quad
    \Sigma_{(13)}=\{f=0,e=0\}.
\end{equation}
For the open stratum, the analysis of Section~\ref{UV-theory} identifies its topology with
\begin{equation}
    \Sigma_e\simeq \left(\SO(3,\mathbb{C})/\SO(3,\mathbb{R})\times U(1)^{2}\right)/\left(\mathbb{Z}_2\times \mathbb{Z}_2\right)
    \simeq \left(\mathbb{R}^3\times U(1)^{2}\right)/\left(\mathbb{Z}_2\times \mathbb{Z}_2\right).
\end{equation}
On the strata $\Sigma_{(12)}$ and $\Sigma_{(23)}$ one can parameterize $S$ as follows
\begin{equation}
    S_{(12)}=
    \begin{pmatrix}
      0&b&c\\
      b&-\bar{c}^2&\bar{b}\bar{c}\\
      c&\bar{b}\bar{c}&-\bar{b}^2
    \end{pmatrix},\quad
      |b|^2+|c|^2=1,\, b\neq0,\quad
          S_{(23)}=
    \begin{pmatrix}
      -\bar{e}^2&\bar{e}\bar{c}&c\\
      \bar{e}\bar{c}&-\bar{c}^2&e\\
      c&e&0
    \end{pmatrix},\quad
      |e|^2+|c|^2=1,\, e\neq0.
\end{equation}
Thus, topologically $\Sigma_{(12)}$ and $\Sigma_{(23)}$ are
\begin{equation}
    \Sigma_{(12)}\simeq S^3\setminus S^1,\quad
    \Sigma_{(23)}\simeq S^3\setminus S^1.
\end{equation}
For the stratum $\Sigma_{(13)}$ one takes
\begin{equation}
S_{(13)}=\begin{pmatrix}
          0&0&e^{i\phi}\\
          0&-e^{-2i\phi}&0\\
          e^{i\phi}&0&0
         \end{pmatrix},\quad\phi\in[0,2\pi).
\end{equation}
Thus,
\begin{equation}
    \Sigma_{(13)}\simeq S^1. 
\end{equation}

Proceeding in the same way, one can argue that the codimension-two strata of $\SU(N)/\SO(N)$ are indexed by adjacent transpositions $\Sigma_{(1,2)}$, $\Sigma_{(2,3)}$, $\Sigma_{(3,4)}$ etc. Moreover, all of these strata are connected for $N>2$. 

It is interesting to see how these codimension two strata are glued to the main stratum $\Sigma_e$ in coordinates of $\SO(N,\mathbb{C})/\SO(N,\mathbb{R})$.

For $N=2$ using the parametrization \eqref{SU(2)-parametrization}, one finds
\begin{equation}
    S=gg^T=
    \begin{pmatrix}
    \frac{e^{i\varphi}}{\cosh\rho}&i \tanh\rho\\
    i\tanh\rho&\frac{e^{-i\varphi}}{\cosh\rho}
    \end{pmatrix}\implies
    \Sigma_e=\{\rho\in(-\infty,\infty),
    \quad\varphi\in[0,2\pi)\},\quad
    \Sigma_{(12)}^{\pm}=\{\rho=\pm\infty\}.
\end{equation}

For $N=3$ it is more convenient to work in Poincar\'e ball coordinates on $\mathbb{H}_3^+$
\begin{equation}
\mathbf y=(y_1,y_2,y_3),\qquad
|\mathbf y|^2<1.
\end{equation}
as
\begin{equation}\label{Poincare-ball-coordinates}
y_1=
\frac{\sinh\rho\,\cos\phi}
{1+\cosh\rho\cosh\tau},\quad
y_2= \frac{\sinh\rho\,\sin\phi}
{1+\cosh\rho\cosh\tau},\quad
y_3=
\frac{\cosh\rho\,\sinh\tau}
{1+\cosh\rho\cosh\tau}.
\end{equation}
In these coordinates, the metric is
\begin{equation}
ds^2=\frac{4\left(dy_1^2+dy_2^2+dy_3^2\right)}
{(1-y_1^2-y_2^2-y_3^2)^2}.
\end{equation}
Using the parameterization \eqref{SU(3)-coset-parametrization}, one finds $S=gg^T$ which has the form  \eqref{S-in-N-3} with
\begin{equation}
    a=\frac{e^{2 i \varphi_1}}{\cosh^2\rho\cosh2\tau+\sinh ^2\rho\cos2\phi},\quad
    f=\frac{e^{-2i\varphi_2}}{\cosh2\rho}.
\end{equation}
Thus, the main stratum $\Sigma_e$ corresponds to the domain of coordinates \eqref{H3-ranges}. The action \eqref{Z2-Z2-action} of the discrete group $\mathbb{Z}_2\times \mathbb{Z}_2$ in Poincar\'e ball coordinates \eqref{Poincare-ball-coordinates} is as follows
\begin{equation}
    \begin{aligned}
        &s_1:\quad \varphi_1\rightarrow \varphi_1+\pi,\quad
        y_1\rightarrow-y_1,\quad y_2\rightarrow-y_2,\\
        &s_2:\quad \varphi_2\rightarrow \varphi_2+\pi,\quad
        y_1\rightarrow-y_1,\quad y_3\rightarrow-y_3.
    \end{aligned}
\end{equation}

For the $a=0$ stratum $\Sigma_{(12)}$, we take  the north pole limit
\begin{equation}
\tau\to+\infty,
\end{equation}
with $\rho$ finite. In Poincar\'e ball coordinates, it corresponds to $(y_1,y_2,y_3)\to (0,0,1)$. Then
\begin{equation}
a\to0,\qquad f=\frac{e^{-2i\varphi_2}}{\cosh2\rho}\neq0.
\end{equation}
Introduce
\begin{equation}
\eta=
\sqrt2\,\sinh\rho\;
e^{i(\varphi_1+\phi-\varphi_2)},
\end{equation}
then the limiting symmetric matrix becomes
\begin{equation}
S_{(12)}=
\begin{pmatrix}
0&
\dfrac{i e^{i\varphi_2}}{\sqrt{1+|\eta|^2}}&
-\dfrac{i\eta}{\sqrt{1+|\eta|^2}}
\\[3mm]
\dfrac{i e^{i\varphi_2}}{\sqrt{1+|\eta|^2}}&
\dfrac{\bar\eta^2}{1+|\eta|^2}&
\dfrac{e^{-i\varphi_2}\bar\eta}{1+|\eta|^2}
\\[3mm]
-\dfrac{i\eta}{\sqrt{1+|\eta|^2}}&
\dfrac{e^{-i\varphi_2}\bar\eta}{1+|\eta|^2}&
\dfrac{e^{-2i\varphi_2}}{1+|\eta|^2}
\end{pmatrix}.
\end{equation}
The limit $\tau\to-\infty$, which is $(y_1,y_2,y_3)\to (0,0,-1)$ in Poincar\'e ball coordinates, gives the same stratum in the quotient
\begin{equation}
s_2:\quad(0,0,1)\leftrightarrow(0,0,-1).
\end{equation}

For the stratum $\Sigma_{(23)}$ we need $f=0$, but $a\neq0$. It corresponds to
\begin{equation}
\tau=\frac{x}{\sqrt{\cosh2\rho}},\qquad
\phi=\pm\frac{\pi}{2}+\frac{y}{\sqrt{\cosh2\rho}},\quad
\rho\to\infty,
\end{equation}
or
\begin{equation}
(y_1,y_2,y_3)\to(0,\pm1,0).
\end{equation}
The limiting matrix is
\begin{equation}
S_{(23)}=
\begin{pmatrix}
\dfrac{e^{2i\varphi_1}}{1+|\xi|^2}&
\dfrac{i\bar\xi}{1+|\xi|^2}&
\dfrac{e^{i\varphi_1}\xi}{\sqrt{1+|\xi|^2}}
\\[3mm]
\dfrac{i\bar\xi}{1+|\xi|^2}&
-\dfrac{e^{-2i\varphi_1}\bar\xi^2}{1+|\xi|^2}&
-\dfrac{i e^{-i\varphi_1}}{\sqrt{1+|\xi|^2}}
\\[3mm]
\dfrac{e^{i\varphi_1}\xi}{\sqrt{1+|\xi|^2}}&
-\dfrac{i e^{-i\varphi_1}}{\sqrt{1+|\xi|^2}}&
0
\end{pmatrix},
\end{equation}
where
\begin{equation}
\xi=e^{-i\varphi_2}(x+iy).
\end{equation}

For the stratum $\Sigma_{(13)}$ one can take either
$|\eta|\to\infty$ in $\Sigma_{(12)}$, or $|\xi|\to\infty$ in $\Sigma_{(23)}$.
Both give
\begin{equation}
S_{(13)}=
\begin{pmatrix}
0&0&e^{i\alpha}\\
0&-e^{-2i\alpha}&0\\
e^{i\alpha}&0&0
\end{pmatrix},
\qquad
\alpha\in S^1.
\end{equation}
\section{Generalization to symmetric spaces of maximal rank}\label{generalizations}
A particularly important class of compact symmetric spaces is formed by symmetric spaces of maximal rank. Let $G/K$ be a compact symmetric space\footnote{We write $K$ for the isotropy subgroup in this section, reserving $H$ for the positive Hermitian field used earlier.}. As previously in \eqref{z2-grading}, we have the decomposition $\mathfrak{g} = \mathfrak{g}^{(0)} \oplus \mathfrak{g}^{(1)}$.

The \textit{rank} of a symmetric space $G/K$ is defined as the dimension of a maximal subspace $\mathfrak a\subset\mathfrak{g}^{(1)}$ such that $\mathfrak{a}$ is an abelian subalgebra of $\mathfrak{g}$. Clearly, $\operatorname{rank}(G/K)\leq \operatorname{rank}G$. The symmetric space is said to have maximal rank when
\begin{equation}\label{maximal-rank-definition}
\operatorname{rank}(G/K)=\operatorname{rank}G.
\end{equation}
The classical $\mathrm{ABCD}$ series of maximal rank symmetric spaces are
\begin{equation}\label{symmetric-spaces-maximal-rank}
\frac{\SU(N)}{\SO(N)},\quad
\frac{\SO(2N+1)}{\SO(N)\times \SO(N+1)},\quad
\frac{\mathrm{Sp}(N)}{\mathrm{U}(N)}\quad\text{and}\quad
\frac{\SO(2N)}{\SO(N)\times \SO(N)}.
\end{equation}
Their respective ranks are $N-1$, $N$, $N$ and $N$, which coincide with the ranks of $\SU(N)$, $\SO(2N+1)$, $\mathrm{Sp}(N)$ and $\SO(2N)$ respectively. There are also five exceptional examples
\begin{equation}\label{exceptional-symmetric-spaces-maximal-rank}
\frac{\mathrm{E}_6}{\mathrm{Sp}(4)},\quad
\frac{\mathrm{E}_7}{\SU(8)}, \quad
\frac{\mathrm{E}_8}{\SO(16)},\quad
\frac{\mathrm{F}_4}{\mathrm{Sp}(3)\times \mathrm{Sp}(1)}\quad\text{and}\quad
\frac{\mathrm{G}_2}{\SO(4)},
\end{equation} 
of ranks $6$, $7$, $8$, $4$ and $2$ respectively.

All of these symmetric spaces have a number of common features, because the full Cartan (maximal abelian) subalgebra is available inside the coset space $\mathfrak{g}^{(1)}$. In particular, their Bruhat-type stratifications admit similar descriptions \cite{RichardsonSpringer1990}. The space
$\SU(N)/\SO(N)$ is the $A_{N-1}$ member of the family \eqref{symmetric-spaces-maximal-rank}. In this section, we show that the main results of this paper are also applicable with mild modification for other symmetric spaces from \eqref{symmetric-spaces-maximal-rank} and \eqref{exceptional-symmetric-spaces-maximal-rank}.

We note that all \emph{undeformed} symmetric-space sigma models \eqref{symmetric-spaces-maximal-rank}-\eqref{exceptional-symmetric-spaces-maximal-rank} are classically integrable and one-loop renormalizable, but only a distinguished subset is known to remain integrable after quantization. A well-known sufficient condition for quantum integrability of the undeformed model is simplicity of the subgroup $K$ of $G/K$ \cite{Abdalla:1982yd}. This condition is only sufficient, but not necessary, since there are (at least two) important exceptions to this simple $K$ requirement. In \cite{Evans:2004ur} it was explicitly shown that the cosets
\begin{equation}
\frac{\SO(2N)}{\SO(N)\times \SO(N)}, \qquad \frac{\mathrm{Sp}(2N)}{\mathrm{Sp}(N)\times \mathrm{Sp}(N)}
\end{equation}
exhibit quantum integrability. The former case is precisely the $D_N$ type symmetric space of maximal rank, while the latter  is not of maximal rank.

Suppose that we have a simple complex Lie algebra $\mathfrak{g}^\mathbb{C}$ with its real compact form $\mathfrak{g}\subset\mathfrak{g}^\mathbb{C}$ and a Cartan-Weyl basis $\{h_k\}_{k=1}^{r}$, $\{e_\alpha\}_{\alpha \,\in \,\Delta}$, $[e_\alpha, e_\beta] = N_{\alpha, \, \beta}\,e_{\alpha+\beta}$, $N_{\alpha,\, \beta} = - N_{-\alpha,\, -\beta}$, where $\Delta$ stands for roots. Then we have the standard realization of the compact form:
\begin{equation}\label{compact_form}
    \mathfrak{g} = \mathrm{span}_\mathbb{R}\big\{ih_k,\,e_\alpha - e_{-\alpha},\, i(e_\alpha + e_{-\alpha})\big\}.
\end{equation}
It is easy to check that the subspace $\mathfrak{k} = \mathrm{span}_\mathbb{R}(e_\alpha - e_{-\alpha})$ is actually a (real) subalgebra of $\mathfrak{g}$. Moreover, this subalgebra is clearly stable with respect to the so-called Chevalley involution
\begin{equation}\label{Chevalley-involution}
    e_\alpha \longmapsto -e_{-\alpha}, \quad h_{k} \longmapsto -h_k.
\end{equation}
This provides a structure of a symmetric pair $(\mathfrak{g},\mathfrak{k})$ of maximal rank. At the group level, we denote the involution \eqref{Chevalley-involution} by $\theta$\,\footnote{For $\SL(N,\mathbb{C})/\SO(N,\mathbb{C})$ one has $\theta(g)=(g^T)^{-1}$.}, and its fixed subgroup by
\begin{equation}
 K^{\mathbb C}=\{g\in G^{\mathbb{C}}:\, \theta(g)=g\}.  
\end{equation}
Then the Cartan embedding $G^{\mathbb C}/K^{\mathbb C} \hookrightarrow G^\mathbb{C}$ is
\begin{equation}
    g\longmapsto S=g[\theta(g)]^{-1}.
\end{equation}
This map is well-defined since replacing $g \mapsto gk$ for $k\in K^{\mathbb C}$ leaves $S$ unchanged.

Let $T\subset G$ be the Cartan maximal torus and choose a positive Borel subgroup $B_+=T^{\mathbb C}\cdot N_+\subset G^{\mathbb C}$. At the Lie-algebra level, one has
\begin{equation}
\mathfrak{g}^{\mathbb C} =\mathfrak{b}_+\oplus\mathfrak k^{\mathbb C}.
\end{equation}
Similarly to the $\SU(N)/\SO(N)$ case considered above, the orbit of the $B_+$  action 
\begin{equation}
S\longmapsto b\,S\,[\theta(b)]^{-1},\quad b\in B_+,
\end{equation}
of the generic point in $G^{\mathbb C}/K^{\mathbb C}$ is open \cite{RichardsonSpringer1990}. We denote its intersection with $G/K$ by $\Sigma_0$. The corresponding residual group
\begin{equation}
\Gamma=B_+\cap K^{\mathbb C}
\end{equation}
is finite. Since the involution $\theta:G^\mathbb{C}\longrightarrow G^\mathbb{C}$ exchanges $B_+$ and $B_-$, this finite group $\Gamma$, in fact, lies in the intersection of $B_+\cap K^\mathbb{C}$ with $T^{\mathbb C}$. The explicit form of this finite group $\Gamma$ depends on a particular symmetric pair.

Similarly to \eqref{g-coset-parametrization}, every point of the principal stratum $\Sigma_0$ admits a representative
\begin{equation}
g=t\Omega V^{-1},
\qquad
t\in T,\quad \Omega\in A\cdot N_+,\quad V\in K^{\mathbb C},
\end{equation}
where $A$ is the positive-real part of $T^{\mathbb C}$. Since both $g$ and $t$ are unitary, one has
\begin{equation}
H=V^\dagger V=\Omega^\dagger\Omega=H^{\dagger}.
\end{equation}
Thus, $H$ belongs to the set of positive Hermitian elements
\begin{equation}
K^{\mathbb C}/K
 =\big\{H\in K^{\mathbb C}:\quad H^\dagger=H,\ H>0\big\}.
\end{equation}

Conversely, given $t\in T$ and $H\in K^{\mathbb C}/K$, take $V=H^{1/2}$ and write its Iwasawa decomposition as $V=k\Omega$, with $k\in G$ and $\Omega\in AN_+$. The corresponding coset representative is $g=tk^{-1}$. The Iwasawa factors are not determined from this decomposition uniquely. The ambiguity/freedom is precisely the action of our finite group $\Gamma$:
\begin{equation}
(t,H)\sim(tD,D^{-1}HD),\qquad D\in\Gamma,
\end{equation}
and therefore
\begin{equation}\label{Sigma-e-generic}
\Sigma_0\simeq (T\times K^{\mathbb C}/K)/\Gamma .
\end{equation}

The open-stratum parametrization \eqref{Sigma-e-generic} suggests a possible candidate for the ultraviolet description found for $\SU(N)/\SO(N)$
\begin{equation}
\bigl(K^{\mathbb C}/K\bigr)\ \mathrm{WZW}\ 
\otimes\ \mathrm{U}(1)^{\operatorname{rank}G}.
\end{equation}
This proposal concerns the ultraviolet fixed point and should not be read as a claim of quantum integrability for every pair in \eqref{symmetric-spaces-maximal-rank} and \eqref{exceptional-symmetric-spaces-maximal-rank}. 

For those deformed sigma models that remain quantum integrable, we conjecture that the screening field \eqref{screening-proposal} admits a natural generalization. The candidate for a WZW primary is the same as for the symmetric-space sine-Gordon models \cite{CastroAlvaredo:2000kq} and corresponds to $\mathfrak{g}^{(1)}_{\mathbb C}$ in the Lie algebra decomposition
\begin{equation}
\mathfrak g_{\mathbb C}
=\mathfrak{g}^{(0)}_{\mathbb C}\oplus
\mathfrak{g}^{(1)}_{\mathbb C}.
\end{equation}
We intend to make the statement more precise elsewhere.
\section{Concluding remarks}\label{concl}
In this paper, we have studied the ultraviolet limit of the Yang--Baxter deformed $\SU(N)/\SO(N)$ sigma model using a complex orthogonal triangularization of a generic coset representative. On the open stratum $\Sigma_e$, the resulting coordinates consist of a compact Cartan torus and a positive Hermitian element of $\SO(N,\mathbb C)$, subject to a finite $(\mathbb Z_2)^{N-1}$ identification. In the leading order (i.e. exactly at the UV fixed point) the metric and $B$-field take the form of that of the $\SO(N,\mathbb C)/\SO(N,\mathbb R)$ WZW background together with $N-1$ free compact bosons. The first subleading correction organizes into a perturbation by a symmetric-traceless WZW primary dressed by Cartan and affine currents. We have identified the chiral part of this perturbation with the screening charge. In the $N=3$ case we have found a nontrivial spin-three density whose integral commutes with this charge, providing a concrete test of the proposed quantum integrable structure.

We also propose a generalization of our results for a wide class of symmetric spaces of maximal rank.

There are several open questions:
\begin{enumerate}
\item It is interesting to express the full deformed metric and
$B$-field \eqref{deformed-g-B}, beyond first orders, directly in terms of the  UV fields $H$ and $\Lambda$. In particular, a direct verification of renormalization group flow equations \eqref{RG-equation} would be valuable.
\item The spin-three conserved density \eqref{IM-invariant-3} should not be special to $N=3$. We have performed an explicit check for $N=4$ and it gives the following integral of motion at the same value of the quantization parameter $\gamma=\pm\frac{1}{\sqrt{k-4}}$
\begin{multline}\label{IM-invariant-4}
G_3=-\frac{4k}{3}\Tr\left(\partial\Lambda^3\right)
-\Tr\left(\partial\Lambda\mathcal{R}(J)^2\right)-\frac{i\gamma}{18}\Tr\left(\mathcal{R}(J)^3\right)+\\+
    \frac{2\gamma k}{3\sqrt{3}}\partial\varphi_1\partial^2\varphi_2+
    \frac{\sqrt{2}\gamma k}{3}\partial\varphi_2\partial^2\varphi_3+
    \frac{\sqrt{2}\gamma k}{3\sqrt{3}}\partial\varphi_1\partial^2\varphi_3.
\end{multline}
We believe results \eqref{IM-invariant-3} and \eqref{IM-invariant-4} to be strong evidence for the existence of an invariant construction for general $N$. 

Comparing \eqref{IM-invariant-3} and \eqref{IM-invariant-4}, we note that the last terms can be expressed as
\begin{equation}
    \mathcal{A}(\partial\Lambda,\partial^2\Lambda)=\sum_{i<j}\left(\partial\Lambda_{ii}\partial^2\Lambda_{jj}-
    \partial\Lambda_{jj}\partial^2\Lambda_{ii}\right),
\end{equation}
modulo total derivatives. Thus, we propose the following ansatz for the conserved density
\begin{equation}\label{IM-invariant-N}
G_3=\xi_1(N,k)\Tr\left(\partial\Lambda^3\right)
-\Tr\left(\partial\Lambda\mathcal{R}(J)^2\right)+\xi_2(N,k)\Tr\left(\mathcal{R}(J)^3\right)+\xi_3(N,k)\mathcal{A}(\partial\Lambda,\partial^2\Lambda).
\end{equation}
\item The integral of motion $I_2=\oint G_3(z)dz$ should belong to an intersection of certain $W$-algebras similarly to the case of $\SO(N+1)/\SO(N)$ \cite{Fateev:2017mug,Fateev:2018yos,Litvinov:2018bou}. This amounts to splitting the screening charge \eqref{screening-proposal} into pieces in such a way that the commutant of each individual piece is some current (not just its zero mode).  There is a natural splitting related to the action of $\SO(N,\mathbb C)$. The positive Hermitian representative of $\SO(N,\mathbb C)$ acts on the spin-two field $\Phi_X^{(2)}$ by $X\mapsto HXH^T$ and in a similar way on WZW currents $J_{ij}$.

Consider for example $N=3$ and the boost $H_\lambda=\exp(i\lambda\,t_{12})$. Define
\begin{equation}
J_\pm=\frac{J_{13}\mp i J_{23}}{2},\qquad
M_{\pm2}=\frac{d_1\mp i s_{12}}2,\qquad
M_{\pm1}=\frac{s_{13}\mp i s_{23}}2.
\end{equation}
If the boost $H_\lambda$ acts on both $\Phi^{(2)}$ and WZW currents, the screening \eqref{screening-proposal} takes the exact form
\begin{equation}\label{Spm-screenings}
S_{\lambda}=e^{2\lambda}S_++\int \partial\varphi_2\Phi_{d_2}dz+e^{-2\lambda}S_-,
\qquad
S_\pm=\int
\left(\partial\varphi_1\pm i\gamma J_{12}\right)\Phi_{M_{\pm2}}dz,
\end{equation}
modulo total derivatives. Thus, $e^{-2\lambda}S_{\lambda}\to S_+$ as $\lambda\to\infty$ and $e^{2\lambda}S_{\lambda}\to S_-$ as $\lambda\to-\infty$. The corresponding large-boost components of $G_3$ are
\begin{equation}\label{W3-proposal}
W_3^{\pm}=(\partial\varphi_1\pm i\gamma J_{12})J_\pm^2.
\end{equation}
The currents \eqref{W3-proposal} commute with the corresponding limiting screening charges $S_{\pm}$ \eqref{Spm-screenings}. The discrete group $\mathbb{Z}_2^2$ permutes two screenings $S_+$ and $S_-$. Boosts in the other coordinate planes give analogous decompositions.
\item Another interesting issue is whether the screening \eqref{screening-proposal}  has an equivalent description by fermionic exponential
screenings, similar to \cite{Fateev:1996ea,Fateev:2017mug,Fateev:2018yos,Litvinov:2018bou,Fateev:2019xuq,Litvinov:2019rlv,Alfimov:2020jpy}.  If such a dual fermionic description exists, it should provide a tractable starting point for computing the two-particle $S$-matrix and comparing it with the expected Yang--Baxter-deformed scattering theory.
\end{enumerate}
We will address these questions elsewhere.
\section*{Acknowledgments}
We acknowledge discussions with Mikhail Bershtein, Boris Feigin, Ben Hoare, Andrew Kuzovchikov and Alexey Rosly. This work was supported by the Russian Science Foundation under grant 26-11-00342. The work of D.A. and A.L. was supported by the BASIS Foundation.
\Appendix
\section{Proof of relations (\ref{relation-1}) and (\ref{relation-2})}\label{relations-proof}

In order to prove \eqref{relation-1} we note that 
\begin{equation}\label{adjoint-of-cartan}
    \textrm{Cartan}\left(
    \Omega\,t\,\Omega^{-1}\right)=t\quad
    \text{for}\quad t\in\textrm{Cartan}(\mathfrak{su}(N)),
\end{equation}
so the matrix $\left(\Omega\,t\,\Omega^{-1}-t\right)$ is strictly upper-triangular. This establishes \eqref{relation-1}:
\begin{equation}
    \left(1+i\mathcal{R}_{\Omega}\right)
\left(t\right)=t+i\Omega^{-1}R\left(
    t+\left(\Omega\,t\,\Omega^{-1}-t\right)\right)\Omega=t+i\Omega^{-1}\left(0+i\left(\Omega\,t\,\Omega^{-1}-t\right)\right)\Omega=\Omega^{-1}t\Omega.
\end{equation}

In order to prove \eqref{relation-2}, one should express $H^{-1}dH$ in terms of $\Omega$
\begin{equation} \label{relation-for-H}
H^{-1}dH=\Omega^{-1}\left({\left(\Omega^{\dagger}\right)}^{-1}d\Omega^{\dagger}\right)\Omega+\Omega^{-1}d\Omega
\end{equation}
and act on both sides of the \eqref{relation-2} with  $\textrm{Ad}\,\Omega$. One arrives at the following equation:
\begin{equation}\label{relation-2-2}
     \left(1+i\mathcal{R}\right)\mathcal{R}_{\Omega^{-1}} \left({\left(\Omega^{\dagger}\right)}^{-1}d\Omega^{\dagger}\right)=-2i{\left(\Omega^{\dagger}\right)}^{-1}d\Omega^{\dagger}.
\end{equation}
It is easy to see that both the \textit{l.h.s.} and the \textit{r.h.s.} of \eqref{relation-2-2} are lower-triangular matrices. Thus, it is sufficient to prove that these matrices have equal strictly lower-triangular and Cartan parts. 

For the first clause, we notice that lower-triangular part of the left-hand side is equal to the lower-triangular part of 
\begin{equation}
    2\mathcal{R}_{\Omega^{-1}} \left({\left(\Omega^{\dagger}\right)}^{-1}d\Omega^{\dagger}\right).
\end{equation}
But the matrix 
\begin{equation}
    \left(2i+2\mathcal{R}_{\Omega^{-1}} \right)\left({\left(\Omega^{\dagger}\right)}^{-1}d\Omega^{\dagger}\right)=2i\left(1-i\mathcal{R}_{\Omega^{-1}} \right)\left({\left(\Omega^{\dagger}\right)}^{-1}d\Omega^{\dagger}\right)
\end{equation}
is clearly upper-triangular. 

The second clause is equivalent to
\begin{equation} \label{relation-2-2-2}
    \textrm{Cartan}\left(\mathcal{R}_{\Omega^{-1}} \left({\left(\Omega^{\dagger}\right)}^{-1}d\Omega^{\dagger}\right)\right)= -2i\,\textrm{Cartan}\left({\left(\Omega^{\dagger}\right)}^{-1}d\Omega^{\dagger}\right).
\end{equation}
The \textit{l.h.s.} can be rewritten as
\begin{equation}
    \textrm{Cartan}\left(\mathcal{R}_{\Omega^{-1}} \left({\left(\Omega^{\dagger}\right)}^{-1}d\Omega^{\dagger}\right)\right)=\textrm{Cartan}\left(\Omega\mathcal{R} \left(\Omega^{-1}\left({\left(\Omega^{\dagger}\right)}^{-1}d\Omega^{\dagger}\right)\Omega\right)\Omega^{-1}\right)=\textrm{Cartan}\left(\Omega\left(i n_{+}-i n_{-}\right)\Omega^{-1}\right).
\end{equation}
Here we have introduced the decomposition 
\begin{equation}
\Omega^{-1}\left({\left(\Omega^{\dagger}\right)}^{-1}d\Omega^{\dagger}\right)\Omega=h+n_{+}+n_{-},
\end{equation}
where $h$ is Cartan, $n_{+}$ is strictly upper-triangular and $n_{-}$ is strictly lower-triangular. Since $\Omega n_{+}\Omega^{-1}$ is strictly upper-triangular and does not contribute to the Cartan part, the equation can be subsequently simplified as
\begin{equation}
    \textrm{Cartan}\left(\Omega\left(i n_{+}-i n_{-}\right)\Omega^{-1}\right)=-i\,\textrm{Cartan}\left(\Omega (h+n_{+}+n_{-})\Omega^{-1}\right)+ih=-i\textrm{Cartan}\left({\left(\Omega^{\dagger}\right)}^{-1}d\Omega^{\dagger}\right)+ih,
\end{equation}
where we have used \eqref{adjoint-of-cartan}. Clearly, \begin{equation}
    \textrm {Cartan}\,(H^{-1}dH)=0.
\end{equation}
Therefore, from \eqref{relation-for-H} one concludes that
\begin{equation}
    h=-\textrm{Cartan}\left(\Omega^{-1} d\Omega\right).
\end{equation}
Moreover, matrix $\Omega$ has real diagonal entries, hence
\begin{equation}
    \textrm{Cartan}\left(\Omega^{-1} d\Omega\right) = \textrm{Cartan}\left({\left(\Omega^{\dagger}\right)}^{-1}d\Omega^{\dagger}\right)
\end{equation}
and 
\begin{equation}
    h=-\textrm{Cartan}\left({\left(\Omega^{\dagger}\right)}^{-1}d\Omega^{\dagger}\right).
\end{equation}
Thus, the second clause of \eqref{relation-2-2} and, consequently, equation \eqref{relation-2} are proven.
\section{Symmetric analog of Bruhat decomposition}\label{symmetric-Bruhat-proof}
In this appendix, following \cite{Gelfand-Naimark}, we prove a symmetric analog of Bruhat decomposition \eqref{symmetric-Bruhat}. Let $E_{ij}$ denote the matrix unit with a $1$ in position $(i,j)$.  If
$i<j$, then
\begin{equation}\label{special-b}
  b=\mathbf{1}+tE_{ij}\in B_+.
\end{equation}
The congruence transformation $S\longmapsto bSb^{T}$ by \eqref{special-b} simultaneously acts as follows
\begin{equation}
  R_i\longmapsto R_i+tR_j,
  \qquad
  C_i\longmapsto C_i+tC_j.
\end{equation}
Here the arrow means replacement: row $i$ and column $i$ are replaced by the corresponding expressions. Explicitly 
\begin{equation}\label{elementary-b-applied-to-S}
\begin{aligned}
  (\mathbf{1}+tE_{ij})S(\mathbf{1}+tE_{ji})
  &=S+tE_{ij}S+tSE_{ji}+t^2E_{ij}SE_{ji}.
\end{aligned}
\end{equation}
We note that in \eqref{elementary-b-applied-to-S} there is the quadratic term. It is produced by making the row and column replacements. A diagonal matrix in $B_+$ also allows one to replace
\begin{equation}
  R_i\longmapsto \lambda R_i,
  \qquad
  C_i\longmapsto \lambda C_i,
  \qquad \lambda\ne0.
\end{equation}

Now we present the algorithm for reducing the symmetric invertible matrix $S=S^T\in \mathrm{GL}(N,\mathbb{C})$ to an involution $P_{\sigma}$. There are two cases
\begin{enumerate}
\item Suppose that $s_{NN}\ne0$.  For every $i<N$, apply the elementary upper triangular matrix
\begin{equation}
  b=\mathbf{1}-\frac{s_{iN}}{s_{NN}}E_{iN}.
\end{equation}
The corresponding  operation eliminates both $s_{iN}$ and $s_{Ni}$. We obtain
\begin{equation}
  S\rightarrow
  \begin{pmatrix}
    S'&0\\
    0&s_{NN}
  \end{pmatrix},
  \qquad S'^{T}=S'.
\end{equation}
Since $S$ is invertible, $S'$ is invertible. Scaling the row and column $N$ by $(s_{NN})^{-1/2}$ gives
\begin{equation}
  S\rightarrow
  \begin{pmatrix}
    S'&0\\
    0&1
  \end{pmatrix}.
\end{equation}
\item Now suppose that $s_{NN}=0$.  Since $S$ is invertible, its last row and column are not zero. Let
\begin{equation}
  r=\max\{i<N:s_{iN}\ne0\}.
\end{equation}
For each $i<r$, we apply the matrix
\begin{equation}
  b=\mathbf{1}-\frac{s_{iN}}{s_{rN}}E_{ir}\in B_+,
\end{equation}
which eliminates the entries $s_{iN}$ and $s_{Ni}$.  The last row and column now have only two nonzero entries $s_{rN}=s_{Nr}$.

Next, we remove the diagonal entry $s_{rr}$.  Apply the matrix
\begin{equation}
  b=\mathbf{1}-\frac{s_{rr}}{2s_{rN}}E_{rN}\in B_+,
\end{equation}
which eliminates the entry $(r,r)$, while the entry $s_{rN}$ is unchanged. Scale the row and column $r$ by $s_{rN}^{-1}$.  On the indices $r,N$ we now have the simple matrix
\begin{equation}
  \begin{pmatrix}0&1\\1&0\end{pmatrix}.
\end{equation}

It remains to clear all other entries in the row and column $r$.  For every $k\ne r,N$, we use the matrix\footnote{Here $s_{kr}$ is the entry of a new matrix, obtained on the previous step. In terms of the original matrix it is 
\begin{equation*}
    \frac{s_{kr}}{s_{rN}}-\frac{s_{kN}s_{rr}}{s_{rN}^2}\quad
    \text{for}\quad k<r,\quad
    \frac{s_{kr}}{s_{rN}}\quad\text{for}\quad k>r.
\end{equation*}}
\begin{equation}
  b=\mathbf{1}-s_{kr}E_{kN}\in B_+.
\end{equation}
Since the row and column $N$ have only one non-zero entry $s_{Nr}=s_{rN}=1$, this operation changes
\begin{equation}
  s_{kr}\longmapsto s_{kr}-s_{kr}s_{Nr}=0,
\end{equation}
and eliminates $s_{rk}$ and $s_{kr}$, without affecting the other indices. Thus, we have reduced the original matrix to the block form
\begin{equation}
  S\rightarrow S'\oplus
  \begin{pmatrix}0&1\\1&0\end{pmatrix}.
\end{equation}
It is clear that the resulting matrix $S'$ is symmetric and invertible.
\end{enumerate}
Applying one of these two steps to the new matrix $S'$, we reduce it again. Proceeding in this way, it is clear that the final matrix is some involution $P_{\sigma}$. The diagonal entries correspond to fixed points of  $\sigma$.

The elimination now establishes the existence. To prove the uniqueness, use the rank-control matrices defined in Section \ref{geometry}
\begin{equation}
(P_\sigma)_{ij}
=\rho_{ij}-\rho_{i+1,j}-\rho_{i,j+1}+\rho_{i+1,j+1}.
\end{equation}
Thus the rank-control matrix reconstructs $P_{\sigma}$ uniquely. 
\bibliographystyle{MyStyle.bst}
\bibliography{references.bib}

\end{document}